\pdfoutput=1
\documentclass[12pt,a4paper]{article}
\usepackage{ifthen} % for conditional statements
\newboolean{pdflatex}
\setboolean{pdflatex}{true} % False for eps figures 

\newboolean{articletitles}
\setboolean{articletitles}{true} % False removes titles in references

\newboolean{uprightparticles}
\setboolean{uprightparticles}{false} %True for upright particle symbols

\def\paperauthors{LHCb collaboration} % Leave as is for PAPER, CONF and FIGURE
\def\paperasciititle{Tests of lepton universality using $B^0\to K^0_S l^+ l^-$ and $B^+\to K^{*+} l^+ l^-$ decays} % Set ASCII title here !! MAKE sure it's only ASCII characters !! 
\def\papertitle{Tests of lepton universality using \BdToKSll\ and \BuToKstll\ decays} % Latex formatted title
\def\paperkeywords{{High Energy Physics}, {LHCb}} % Comma separated list
\def\papercopyright{\the\year\ CERN for the benefit of the LHCb collaboration} % new since 9/Apr/2018
\def\paperlicence{CC BY 4.0 licence}
\def\paperlicenceurl{https://creativecommons.org/licenses/by/4.0/}

\usepackage[top=1in, bottom=1.25in, left=1in, right=1in]{geometry}

\usepackage{microtype}
\usepackage{lineno}  % for line numbering during review
\usepackage{xspace} % To avoid problems with missing or double spaces after
\usepackage{caption} %these three command get the figure and table captions automatically small

\usepackage{graphicx}  % to include figures (can also use other packages)
\usepackage{color}
\usepackage{colortbl}
\graphicspath{{./figs/}} % Make Latex search fig subdir for figures
\usepackage{amsmath} % Adds a large collection of math symbols
\usepackage{amssymb}
\usepackage{amsfonts}
\usepackage{upgreek} % Adds in support for greek letters in roman typeset

\newcommand*\patchAmsMathEnvironmentForLineno[1]{%
\expandafter\let\csname old#1\expandafter\endcsname\csname #1\endcsname
\expandafter\let\csname oldend#1\expandafter\endcsname\csname
end#1\endcsname
 \renewenvironment{#1}%
   {\linenomath\csname old#1\endcsname}%
   {\csname oldend#1\endcsname\endlinenomath}%
}
\newcommand*\patchBothAmsMathEnvironmentsForLineno[1]{%
  \patchAmsMathEnvironmentForLineno{#1}%
  \patchAmsMathEnvironmentForLineno{#1*}%
}
\AtBeginDocument{%
\patchBothAmsMathEnvironmentsForLineno{equation}%
\patchBothAmsMathEnvironmentsForLineno{align}%
\patchBothAmsMathEnvironmentsForLineno{flalign}%
\patchBothAmsMathEnvironmentsForLineno{alignat}%
\patchBothAmsMathEnvironmentsForLineno{gather}%
\patchBothAmsMathEnvironmentsForLineno{multline}%
\patchBothAmsMathEnvironmentsForLineno{eqnarray}%
}

\usepackage{hyperxmp}

\usepackage[pdftex,
            pdfauthor={\paperauthors},
            pdftitle={\paperasciititle},
            pdfkeywords={\paperkeywords},
            pdfcopyright={Copyright (C) \papercopyright},
            pdflicenseurl={\paperlicenceurl}]{hyperref}
\usepackage[colorinlistoftodos,textsize=scriptsize]{todonotes}

\usepackage[bottom,flushmargin,hang,multiple]{footmisc}

\usepackage[all]{hypcap} % Internal hyperlinks to floats.

\usepackage{xspace} 
\usepackage{upgreek}

\def\babar  {\mbox{BaBar}\xspace}
\def\belle  {\mbox{Belle}\xspace}

\def\MagUp {\mbox{\em Mag\kern -0.05em Up}\xspace}

\ifthenelse{\boolean{uprightparticles}}%
{

 \def\Pmu         {\ensuremath{\upmu}\xspace}

 \def\Ppi         {\ensuremath{\uppi}\xspace}

 \def\Ppsi        {\ensuremath{\uppsi}\xspace}

 \def\PDelta      {\ensuremath{\Delta}\xspace}                 
 \def\PXi         {\ensuremath{\Xi}\xspace}                 
 \def\PLambda     {\ensuremath{\Lambda}\xspace}                 
 \def\PSigma      {\ensuremath{\Sigma}\xspace}                 
 \def\POmega      {\ensuremath{\Omega}\xspace}                 
 \def\PUpsilon    {\ensuremath{\Upsilon}\xspace}

 \def\PB      {\ensuremath{\mathrm{B}}\xspace}                 
                  
 \def\PD      {\ensuremath{\mathrm{D}}\xspace}

 \def\PJ      {\ensuremath{\mathrm{J}}\xspace}                 
 \def\PK      {\ensuremath{\mathrm{K}}\xspace}

 \def\Pb      {\ensuremath{\mathrm{b}}\xspace}

 \def\Pe      {\ensuremath{\mathrm{e}}\xspace}

 \def\Pi      {\ensuremath{\mathrm{i}}\xspace}

 \def\Pp      {\ensuremath{\mathrm{p}}\xspace}

 \def\Ps      {\ensuremath{\mathrm{s}}\xspace}

 \def\thebaroffset{0.0em}
}
{

 \def\Pmu         {\ensuremath{\mu}\xspace}

 \def\Ppi         {\ensuremath{\pi}\xspace}

 \def\Ppsi        {\ensuremath{\psi}\xspace}                 
                  
 \mathchardef\PDelta="7101
 \mathchardef\PXi="7104
 \mathchardef\PLambda="7103
 \mathchardef\PSigma="7106
 \mathchardef\POmega="710A
 \mathchardef\PUpsilon="7107
                  
 \def\PB      {\ensuremath{B}\xspace}                 
                  
 \def\PD      {\ensuremath{D}\xspace}

 \def\PJ      {\ensuremath{J}\xspace}                 
 \def\PK      {\ensuremath{K}\xspace}

 \def\Pb      {\ensuremath{b}\xspace}

 \def\Pe      {\ensuremath{e}\xspace}

 \def\Pi      {\ensuremath{i}\xspace}

 \def\Pp      {\ensuremath{p}\xspace}

 \def\Ps      {\ensuremath{s}\xspace}

 \def\thebaroffset{0.18em}
}
\newcommand{\offsetoverline}[2][\thebaroffset]{\kern #1\overline{\kern -#1 #2}}%

\makeatletter
\ifcase \@ptsize \relax% 10pt
  \newcommand{\miniscule}{\@setfontsize\miniscule{4}{5}}% \tiny: 5/6
\or% 11pt
  \newcommand{\miniscule}{\@setfontsize\miniscule{5}{6}}% \tiny: 6/7
\or% 12pt
  \newcommand{\miniscule}{\@setfontsize\miniscule{5}{6}}% \tiny: 6/7
\fi
\makeatother

\DeclareRobustCommand{\optbar}[1]{\shortstack{{\miniscule (\rule[.5ex]{1.25em}{.18mm})}
  \\ [-.7ex] $#1$}}

\def\en         {{\ensuremath{\Pe^-}}\xspace}   % electron negative (\em is taken)
\def\ep         {{\ensuremath{\Pe^+}}\xspace}

\def\epem       {{\ensuremath{\Pe^+\Pe^-}}\xspace}

\def\mup        {{\ensuremath{\Pmu^+}}\xspace}
\def\mun        {{\ensuremath{\Pmu^-}}\xspace} % muon negative (\mum is taken)

\def\lepton     {{\ensuremath{\ell}}\xspace}
\def\ellm       {{\ensuremath{\ell^-}}\xspace}
\def\ellp       {{\ensuremath{\ell^+}}\xspace}

\def\squark    {{\ensuremath{\Ps}}\xspace}

\def\bquark    {{\ensuremath{\Pb}}\xspace}

\def\pion   {{\ensuremath{\Ppi}}\xspace}

\def\pip    {{\ensuremath{\pion^+}}\xspace}
\def\pim    {{\ensuremath{\pion^-}}\xspace}

\def\kaon    {{\ensuremath{\PK}}\xspace}
\def\KorKbar {\kern \thebaroffset\optbar{\kern -\thebaroffset \PK}{}\xspace}
\def\Kz      {{\ensuremath{\kaon^0}}\xspace}

\def\Kp      {{\ensuremath{\kaon^+}}\xspace}
\def\Km      {{\ensuremath{\kaon^-}}\xspace}

\def\KS      {{\ensuremath{\kaon^0_{\mathrm{S}}}}\xspace}

\def\Kstarz  {{\ensuremath{\kaon^{*0}}}\xspace}

\def\Kstar   {{\ensuremath{\kaon^*}}\xspace}

\def\Kstarp  {{\ensuremath{\kaon^{*+}}}\xspace}

\def\Dbar    {{\ensuremath{\offsetoverline{\PD}}}\xspace}
\def\D       {{\ensuremath{\PD}}\xspace}

\def\DorDbar {\kern \thebaroffset\optbar{\kern -\thebaroffset \PD}\xspace}

\def\Dzb     {{\ensuremath{\Dbar{}^0}}\xspace}
\def\Dp      {{\ensuremath{\D^+}}\xspace}
\def\Dm      {{\ensuremath{\D^-}}\xspace}

\def\DpDm    {\ensuremath{\Dp {\kern -0.16em \Dm}}\xspace}

\def\B       {{\ensuremath{\PB}}\xspace}

\def\BorBbar {\kern \thebaroffset\optbar{\kern -\thebaroffset \PB}\xspace}
\def\Bz      {{\ensuremath{\B^0}}\xspace}

\def\Bd      {{\ensuremath{\B^0}}\xspace}

\def\BdorBdbar {\kern \thebaroffset\optbar{\kern -\thebaroffset \Bd}\xspace}
\def\Bu      {{\ensuremath{\B^+}}\xspace}

\def\Bp      {{\ensuremath{\Bu}}\xspace}

\def\Bs      {{\ensuremath{\B^0_\squark}}\xspace}

\def\BsorBsbar {\kern \thebaroffset\optbar{\kern -\thebaroffset \Bs}\xspace}

\def\jpsi     {{\ensuremath{{\PJ\mskip -3mu/\mskip -2mu\Ppsi}}}\xspace}
\def\psitwos  {{\ensuremath{\Ppsi{(2S)}}}\xspace}

\def\Y#1S{\ensuremath{\PUpsilon{(#1S)}}\xspace}

\def\proton      {{\ensuremath{\Pp}}\xspace}

\def\Lz          {{\ensuremath{\PLambda}}\xspace}

\def\LorLbar     {\kern \thebaroffset\optbar{\kern -\thebaroffset \PLambda}\xspace}

\def\Lb           {{\ensuremath{\Lz^0_\bquark}}\xspace}

\newcommand{\decay}[2]{\ensuremath{{#1\!\to #2}\xspace}} 

\def\to                 {\ensuremath{\rightarrow}\xspace}

\def\qsq       {{\ensuremath{q^2}}\xspace}

\def\AT#1     {\ensuremath{A_{\mathrm{T}}^{#1}}\xspace}           % 2

\def\C#1      {\ensuremath{\mathcal{C}_{#1}}\xspace}                       % 9
\def\Cp#1     {\ensuremath{\mathcal{C}_{#1}^{'}}\xspace}                    % 7
\def\Ceff#1   {\ensuremath{\mathcal{C}_{#1}^{\mathrm{(eff)}}}\xspace}        % 9  
\def\Cpeff#1  {\ensuremath{\mathcal{C}_{#1}^{'\mathrm{(eff)}}}\xspace}       % 7
\def\Ope#1    {\ensuremath{\mathcal{O}_{#1}}\xspace}                       % 2
\def\Opep#1   {\ensuremath{\mathcal{O}_{#1}^{'}}\xspace}                    % 7

\newcommand{\aunit}[1]{\ensuremath{\text{\,#1}}}       
\newcommand{\tev}{\aunit{Te\kern -0.1em V}\xspace}
\newcommand{\gev}{\aunit{Ge\kern -0.1em V}\xspace}
\newcommand{\mev}{\aunit{Me\kern -0.1em V}\xspace}
\newcommand{\kev}{\aunit{ke\kern -0.1em V}\xspace}
\newcommand{\ev}{\aunit{e\kern -0.1em V}\xspace}
 
\newcommand{\mevc}{\ensuremath{\aunit{Me\kern -0.1em V\!/}c}\xspace}
\newcommand{\gevc}{\ensuremath{\aunit{Ge\kern -0.1em V\!/}c}\xspace}
\newcommand{\mevcc}{\ensuremath{\aunit{Me\kern -0.1em V\!/}c^2}\xspace}
\newcommand{\gevcc}{\ensuremath{\aunit{Ge\kern -0.1em V\!/}c^2}\xspace}
\newcommand{\gevgevcccc}{\ensuremath{\gev^2\!/c^4}\xspace} % for q^2

\def\fb   {\ensuremath{\aunit{fb}}\xspace}
\def\invfb   {\ensuremath{\fb^{-1}}\xspace}

\def\gsim{{~\raise.15em\hbox{$>$}\kern-.85em
          \lower.35em\hbox{$\sim$}~}\xspace}
\def\lsim{{~\raise.15em\hbox{$<$}\kern-.85em
          \lower.35em\hbox{$\sim$}~}\xspace}

\def\pt         {\ensuremath{p_{\mathrm{T}}}\xspace}

\def\photos     {\mbox{\textsc{Photos}}\xspace}

\def\pythia     {\mbox{\textsc{Pythia}}\xspace}

\def\tell1  {TELL1\xspace}
\def\ukl1   {UKL1\xspace}

\def\BdToKSee  {\decay{\Bd}{\KS\epem}}
\def\BdToKSmm  {\decay{\Bd}{\KS\mup\mun}}
\def\BdToKSll  {\decay{\Bd}{\KS \ellp \ellm}}

\def\BdToKzee  {\decay{\Bd}{\Kz\epem}}

\def\BdToKSJpsiee {\decay{\Bd}{\jpsi \left(\epem \right)\KS }}
\def\BdToKSJpsimm {\decay{\Bd}{\jpsi \left(\mup\mun \right)\KS }}

\def\BdToKzJpsiee {\decay{\Bd}{\jpsi \left(\epem \right)\Kz }}

\def\BdToKzJpsi {\decay{\Bd}{\jpsi \Kz }}

\def\BuToKstee  {\decay{\Bp}{\Kstarp\epem}}
\def\BuToKstmm  {\decay{\Bp}{\Kstarp\mup\mun}}
\def\BuToKstll  {\decay{\Bp}{\Kstarp \ellp \ellm}}

\def\BuToKstJpsiee {\decay{\Bu}{\jpsi \left(\epem \right)\Kstarp }}
\def\BuToKstJpsimm {\decay{\Bu}{\jpsi \left(\mup\mun \right)\Kstarp }}
\def\BuToKstJpsill {\decay{\Bu}{\jpsi \left(\ellp \ellm \right)\Kstarp }}
\def\BuToKstJpsi {\decay{\Bu}{\jpsi }\Kstarp }

\def\BuToKstpsitwosll {\decay{\Bu}{\psitwos \left(\ellp\ellm \right)\Kstarp }}

\def\bTos  {\decay{\bquark}{\squark}}
\def\bTosll  {\decay{\bquark}{\squark\ellp\ellm}}
\def\bTosmm  {\decay{\bquark}{\squark\mup\mun}}
\def\bTosee  {\decay{\bquark}{\squark\epem}}

\def\BToHee  {\decay{\B}{H\epem}}
\def\BToHmm  {\decay{\B}{H\mup\mun}}

\def\BToKee  {\decay{\B}{K^{(*)}\epem}}
\def\BToKmm  {\decay{\B}{K^{(*)}\mup\mun}}
\def\BToKll  {\decay{\B}{K^{(*)}\ellp\ellm}}
\def\BToKJpsiee  {\decay{\B}{\jpsi\left(\epem\right)K^{(*)}}}
\def\BToKJpsimm  {\decay{\B}{\jpsi\left(\mup\mun\right)K^{(*)}}}

\def\BToKJpsi  {\decay{\B}{\jpsi K^{(*)}}}
\def\BToKpsitwosee  {\decay{\B}{\psitwos\left(\epem\right)K^{(*)}}}
\def\BToKpsitwosmm  {\decay{\B}{\psitwos\left(\mup\mun\right)K^{(*)}}}

\def\BToKpsitwos  {\decay{\B}{\psitwos K^{(*)}}}

\def\RKS      {\ensuremath{R_{\KS}}\xspace}
\def\RKstp      {\ensuremath{R_{\Kstarp}}\xspace}
\def\RKp      {\ensuremath{R_{\Kp}}\xspace}
\def\RKstz      {\ensuremath{R_{\Kstarz}}\xspace}
\def\RpK      {\ensuremath{R_{\proton \kaon }}\xspace}

\def\RKbrackst      {\ensuremath{R_{\kaon^{(*)}}}\xspace}

\def\RinvKS      {\ensuremath{R^{-1}_{\KS}}\xspace}
\def\RinvKstp      {\ensuremath{R^{-1}_{\Kstarp}}\xspace}

\def\RinvK      {\ensuremath{R^{-1}_{\kaon^{(*)}}}\xspace}

\def\rJpsiK {\ensuremath{r_{\jpsi K^{(*)}}}\xspace}

\def\rinvJpsiK {\ensuremath{r^{-1}_{\jpsi K^{(*)}}}\xspace}
\def\rinvJpsiKS {\ensuremath{r^{-1}_{\jpsi \KS}}\xspace}
\def\rinvJpsiKstp {\ensuremath{r^{-1}_{\jpsi \Kstarp}}\xspace}

\def\RinvpsitwosK {\ensuremath{R^{-1}_{\psitwos K^{(*)}}}\xspace}
\def\RinvpsitwosKS {\ensuremath{R^{-1}_{\psitwos \KS}}\xspace}
\def\RinvpsitwosKstp {\ensuremath{R^{-1}_{\psitwos \Kstarp}}\xspace}

\def\BdToKstll  {\decay{\Bd}{\Kstarz \ellp \ellm}}
\def\BuToKll  {\decay{\Bp}{\Kp \ellp \ellm}}
\def\LbTopKll  {\decay{\Lb}{\proton \Km \ellp \ellm}}
\def\LbToLll  {\decay{\Lb}{\Lz \ellp \ellm}}

\def\BsToKSJpsill  {\decay{\Bs}{\jpsi\left(\ellp\ellm\right)\KS}}
\def\BdToKpimumu {\decay{\Bd}{\Kp \pim \mup \mun}}
\def\BdToXDKSY {\decay{\Bd}{\Dm (\KS X)Y}}
\def\BuToXDKSY {\decay{\Bp}{\Dzb (\KS\pip X)Y}}
\def\BdToKSpipi  {\decay{\Bd}{\KS\pip\pim}}
\def\BuToKstpipi  {\decay{\Bp}{\Kstarp\pip\pim}} % Add in custom defined symbols for this analysis

\usepackage{cite} % Allows for ranges in citations
\usepackage{mciteplus}
\usepackage{longtable} % only for template; not usually to be used in PAPERs

\begin{document}

%%%%%%%%%%%%%%%%%%%%%%%%%
%%%%% Title     %%%%%%%%%
%%%%%%%%%%%%%%%%%%%%%%%%%
\renewcommand{\thefootnote}{\fnsymbol{footnote}}
\setcounter{footnote}{1}

% %%%%%%% CHOOSE TITLE PAGE--------
%\onecolumn
%\input{title-LHCb-INT}
%\input{title-LHCb-ANA}
%\input{title-LHCb-CONF}
%\input{title-LHCb-FIGURE}
% ===============================================================================
% Purpose: LHCb-PAPER journal paper title page template
% Author: 
% Created on: 2010-09-25
% ===============================================================================

%%%%%%%%%%%%%%%%%%%%%%%%%
%%%%%  TITLE PAGE  %%%%%%
%%%%%%%%%%%%%%%%%%%%%%%%%
\begin{titlepage}
\pagenumbering{roman}

% Header ---------------------------------------------------
\vspace*{-1.5cm}
\centerline{\large EUROPEAN ORGANIZATION FOR NUCLEAR RESEARCH (CERN)}
\vspace*{1.5cm}
\noindent
\begin{tabular*}{\linewidth}{lc@{\extracolsep{\fill}}r@{\extracolsep{0pt}}}
\ifthenelse{\boolean{pdflatex}}% Logo format choice
{\vspace*{-1.5cm}\mbox{\!\!\!\includegraphics[width=.14\textwidth]{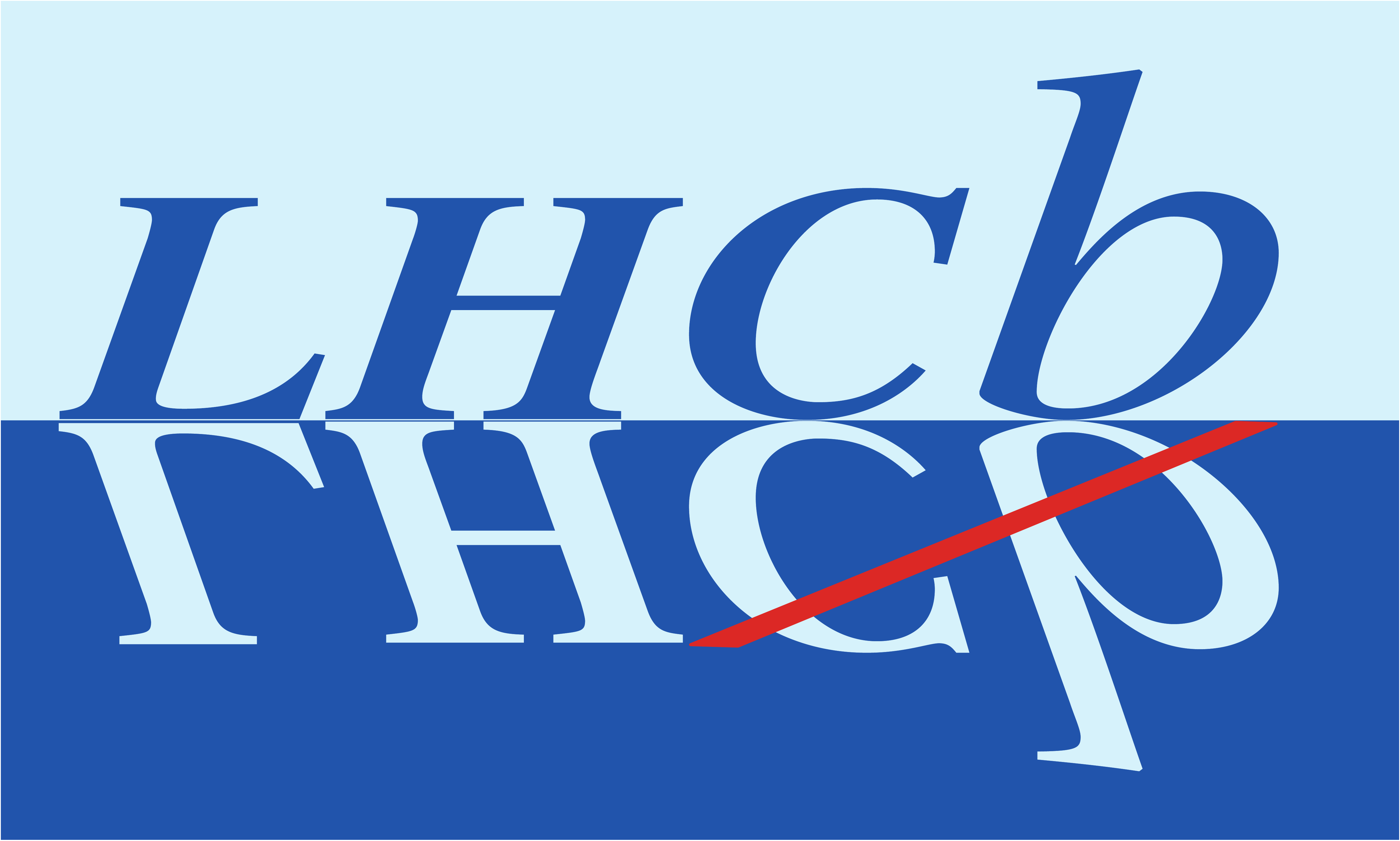}} & &}%
{\vspace*{-1.2cm}\mbox{\!\!\!\includegraphics[width=.12\textwidth]{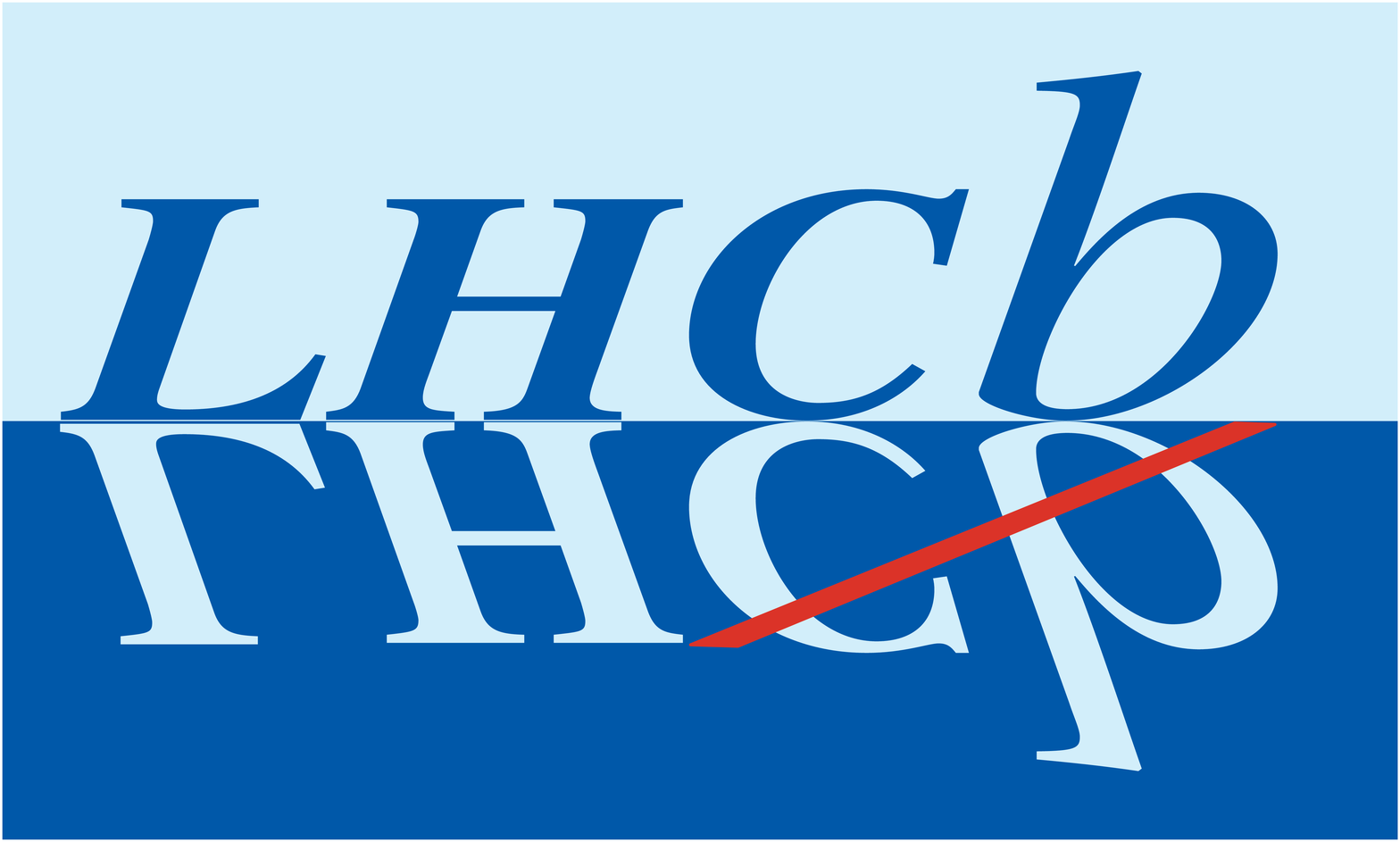}} & &}%
\\
 & & CERN-EP-2021-208 \\  % ID 
 & & LHCb-PAPER-2021-038 \\  % ID 
 & & May 12, 2022 \\ % Date - Can also hardwire e.g.: 23 March 2010
 & & \\
% not in paper \hline
\end{tabular*}

\vspace*{4.0cm}

% Title --------------------------------------------------
{\normalfont\bfseries\boldmath\huge
\begin{center}
% DO NOT EDIT HERE. Instead edit macro in main.tex to keep metadata correct
  \papertitle 
\end{center}
}

\vspace*{2.0cm}

% Authors -------------------------------------------------
\begin{center}
%In the footnote, replace 'paper' by 'Letter' in case of submission to PRL or PLB 
% Edit macro in main.tex to keep metadata correct
\paperauthors\footnote{Authors are listed at the end of this paper.}
\end{center}

\vspace{\fill}

\vspace*{0.3cm}
This paper is dedicated to the memory of our friend and colleague Sheldon Stone.
\vspace*{0.6cm}

% Abstract -----------------------------------------------
\begin{abstract}
  \noindent
  Tests of lepton universality in \BdToKSll\ and \BuToKstll\ decays where \lepton is either an electron or a muon are presented. The differential branching fractions of \BdToKSee\ and \BuToKstee\ decays are measured in intervals of the dilepton invariant mass squared. The measurements are performed using proton-proton collision data recorded by the LHCb experiment, corresponding to an integrated luminosity of 9\invfb. The results are consistent with the Standard Model and previous tests of lepton universality in related decay modes. The first observation of \BdToKSee\ and \BuToKstee\ decays is reported.
  
\end{abstract}

\vspace*{2.0cm}

\begin{center}
 Published in Phys. Rev. Lett. 128, 191802
\end{center}

\vspace{\fill}

{\footnotesize 
% Edit macro in main.tex to keep metadata correct
\centerline{\copyright~\papercopyright. \href{\paperlicenceurl}{\paperlicence}.}}
\vspace*{2mm}

\end{titlepage}

%%%%%%%%%%%%%%%%%%%%%%%%%%%%%%%%
%%%%%  EOD OF TITLE PAGE  %%%%%%
%%%%%%%%%%%%%%%%%%%%%%%%%%%%%%%%

%  empty page follows the title page ----
\newpage
\setcounter{page}{2}
\mbox{~}
%\newpage
%
%% Author List ----------------------------
%%  You need to get a new author list!
%\input{LHCb_authorlist.tex}
%
%The author list for journal publications is provided by the Membership Committee shortly after 'approval to go to paper' has been given.
%%It will be made available on the page
%%\verb!http://www.physik.uzh.ch/~strauman/forMemCo/LHCb-PAPER-XXXX-XXX/! .
%It will be sent to you by email shortly after a paper number has beens assigned.
%The author list should be included already at first circulation, 
%to allow new members of the collaboration to verify whether they have been included correctly.
%Occasionally a misspelled name is corrected or associated institutions become full members.
%In that case, a new author list will be sent to you.
%In case line numbering doesn't work well after including the authorlist, try moving the \verb!\bigskip! after the last author to a separate line.
%
%
%The authorship for Conference Reports should be ``The LHCb
%  collaboration'', with a footnote giving the name(s) of the contact
%  author(s), but without the full list of collaboration names.

% \twocolumn
% %%%%%%%%%%%%% ---------

\renewcommand{\thefootnote}{\arabic{footnote}}
\setcounter{footnote}{0}

%%%%%%%%%%%%%%%%%%%%%%%%%%%%%%%%
%%%%%  Table of Content   %%%%%%
%%%%%%%%%%%%%%%%%%%%%%%%%%%%%%%%
%%%% Uncomment if desired
%\tableofcontents
\cleardoublepage

%%%%%%%%%%%%%%%%%%%%%%%%%
%%%%% Main text %%%%%%%%%
%%%%%%%%%%%%%%%%%%%%%%%%%

\pagestyle{plain} % restore page numbers for the main text
\setcounter{page}{1}
\pagenumbering{arabic}

%% Uncomment during review phase. 
%% Comment before a final submission.
% \linenumbers

%% This is the main body
%% It is useful to have a single file so comemnts are not missed in overleaf.
The \BdToKSll\ and \BuToKstll\ decays, where \lepton refers to either an electron or a muon, are flavour-changing neutral current (FCNC) transitions involving the transformation of a beauty quark into a strange quark.\footnote{Charge conjugate processes are implied throughout.} These decays proceed via higher order electroweak processes in the Standard Model (SM) due to the absence of first order FCNC transitions, making them highly suppressed. Therefore, these decays may receive significant contributions from new quantum fields that lie beyond the Standard Model (BSM) and hence are promising laboratories for new physics (NP) searches.

In recent years, studies of similar \bTosll\ transitions, most prominently \BuToKll\ and \BdToKstll\ decays, have revealed tensions with the SM predictions. Deviations have been seen in ratios of branching fractions
\begin{equation}
    R_{H} \equiv \dfrac{\displaystyle\int^{q^2_{\rm \max}}_{q^2_{\rm \min}} \dfrac{\mathrm{d}\mathcal{B}\left(\BToHmm \right)}{\mathrm{d}\qsq}\mathrm{d}\qsq}{ \displaystyle\int^{q^2_{\rm \max}}_{q^2_{\rm \min}} \dfrac{\mathrm{d}\mathcal{B}\left(\BToHee \right)}{\mathrm{d}\qsq}\mathrm{d}\qsq},
\end{equation}
where $B$ denotes a \Bp or a \Bz meson, $H$ is either a \kaon or a \Kstar meson, and \qsq is the dilepton invariant mass squared. In the SM the charged leptons have identical interaction strengths, which is referred to as lepton universality. The only exception is their interaction with the Higgs field, which determines their differing masses. Therefore, these ratios are predicted to be very close to unity \cite{Hiller:2003js}, with corrections from QED up to $\mathcal{O}(10^{-2})$ \cite{Bordone:2016gaq,Isidori:2020acz} and further small corrections due to the muon-electron mass difference. Furthermore, these ratios benefit from precise cancellation of the hadronic uncertainties that affect predictions of the branching fractions and angular observables, which affect the ratios at $\mathcal{O}(10^{-4})$ \cite{Bobeth:2007dw}. Significant deviation from unity in such ratios would therefore constitute unambiguous evidence of BSM physics.

The ratio \RKstz, measured by the LHCb collaboration using the data collected in the \qsq regions $0.045 < \qsq < 1.1$\gevgevcccc and $1.1 < \qsq < 6.0$\gevgevcccc\cite{LHCb-PAPER-2017-013}, is in tension with the SM predictions at 2.2--2.4 and 2.4--2.5 standard deviations ($\sigma$), respectively, where the ranges are due to the use of different Standard Model predictions. A measurement of \RKp performed in the region $1.1 < \qsq < 6.0$\gevgevcccc deviates from the SM by 3.1 standard deviations \cite{LHCb-PAPER-2021-004}. The analogous ratio measured using \LbTopKll\ decays, \RpK, is consistent with the SM within one standard deviation \cite{LHCb-PAPER-2019-040}. All four measurements show a deficit of \bTosmm\ decays with respect to \bTosee\ decays. 

In addition, angular observables~\cite{LHCb-PAPER-2020-041, LHCb-PAPER-2020-002, LHCb-PAPER-2015-051, LHCb-PAPER-2021-022, ATLAS:2018gqc, BaBar:2006tnv, BaBar:2015wkg, Belle:2009zue, Belle:2016fev, CDF:2011tds, CMS:2015bcy, CMS:2017rzx} and branching fractions~\cite{LHCb-PAPER-2016-012, LHCb-PAPER-2021-014, LHCb-PAPER-2014-006, LHCb-PAPER-2015-009} of \bTosmm\ decays have been measured, with several in tension with the SM. However, the extent to which they may be affected by residual quantum chromodynamics contributions remains uncertain \cite{Descotes-Genon:2015uva, Jager:2014rwa, Lyon:2014hpa, Khodjamirian:2012rm, Khodjamirian:2010vf, Descotes-Genon:2014uoa, Horgan:2013pva, Beaujean:2013soa, Hambrock:2013zya, Altmannshofer:2013foa, Bobeth:2017vxj}.

Intriguingly, it is possible to account for all these anomalies simultaneously through the modification of the \bTos\ coupling in a model-independent way \cite{Ciuchini:2020gvn, Kowalska:2019ley, Alguero:2019ptt, Hurth:2020rzx, Ciuchini:2019usw, Aebischer:2019mlg, Alok:2019ufo, Altmannshofer:2021qrr, Geng:2021nhg, Cornella:2021sby, Alguero:2021anc, Hurth:2021nsi}. Such a modification can be generated by the presence of a heavy neutral boson~\cite{Altmannshofer:2014cfa,Crivellin:2015mga,Celis:2015ara,Falkowski:2015zwa,Allanach:2019mfl,Allanach:2019iiy,Kawamura:2019rth,Dwivedi:2019uqd,Han:2019diw,Capdevila:2020rrl,Altmannshofer:2019xda,Chen:2020szf,Carvunis:2020exc,Karozas:2020zvv,Borah:2020swo,Allanach:2020kss,Sheng:2021tom} or a leptoquark~\cite{Hiller:2014yaa,DiLuzio:2017vat,Greljo:2018tuh,Gripaios:2014tna,deMedeirosVarzielas:2015yxm,Barbieri:2016las,Bordone:2018nbg,Fornal:2018dqn,Balaji:2019kwe,Cornella:2019hct,Datta:2019tuj,Popov:2019tyc,Bigaran:2019bqv,Bernigaud:2019bfy,DaRold:2019fiw,Fuentes-Martin:2019bue,Hati:2019ufv,Datta:2019bzu,Crivellin:2019dwb,Borschensky:2020hot,Saad:2020ihm,Fuentes-Martin:2020bnh,BhupalDev:2020zcy,Fornal:2020ngq,Davighi:2020qqa,Babu:2020hun, Angelescu:2021lln}, as well as in models with supersymmetry~\cite{Trifinopoulos:2019lyo,Hu:2019ahp,Hu:2020yvs}, extra dimensions~\cite{Shaw:2019fin}, and extended Higgs sectors~\cite{Barman:2018jhz,Arnan:2019uhr,DelleRose:2019ukt,Ordell:2019zws,Marzo:2019ldg}.

The \BdToKSll\ and \BuToKstll\ decays are the isospin partners of \BuToKll\ and \BdToKstll\ decays and are expected to be affected by the same NP contributions. Testing lepton universality by measuring the ratios \RKS and \RKstp can therefore provide important additional evidence for or against NP. However, while these decays have similar branching fractions to their isospin partners, $\mathcal{O}(10^{-6})$ to $\mathcal{O}(10^{-7})$, they suffer from a reduced experimental efficiency at LHCb due to the presence of a long-lived \KS meson in the final state. These ratios have previously been measured by the \babar \cite{BaBar:2012mrf} and \belle \cite{Belle:2019oag, BELLE:2019xld} collaborations. The differential branching fractions of the muon modes, \BdToKSmm\ and \BuToKstmm, were found to be lower although still consistent with predictions at low \qsq in a measurement performed by the LHCb collaboration \cite{LHCb-PAPER-2014-006}. No single experiment has unambiguously observed the electron decay modes to date. 

In this Letter, measurements of the ratios \RKS and \RKstp and the differential branching fractions of \BdToKSee\ and \BuToKstee\ decays are presented. The measurements are performed using proton-proton ($pp$) collision data corresponding to an integrated luminosity of 9\invfb recorded by the LHCb experiment in 2011, 2012 (Run 1) and 2016--2018 (Run 2) at centre-of-mass energies of 7, 8 and 13\tev, respectively. The \KS and \Kstarp mesons are reconstructed in the \pip\pim and \KS\pip  final states, respectively. The ratio \RKS and the branching fraction $\mathcal{B}(\BdToKSee)$ are measured in the region $1.1 < \qsq < 6.0$\gevgevcccc, while \RKstp and $\mathcal{B}(\BuToKstee)$ are determined in the range $0.045 < \qsq < 6.0$\gevgevcccc. A wider range is used in the case of the \Bp decay, the differential branching fraction of which is enhanced at low \qsq by the photon pole, since the \Kstarp is a vector meson. Splitting the \qsq range into two bins at 1.1 \gevgevcccc, as was done in the \RKstz measurement, is not possible due to the limited data sample.

The analysis is designed to minimise systematic uncertainties, particularly those associated with differences in the detector response between electrons and muons. The ratios and differential branching fractions are normalised to the control modes, \BdToKSJpsiee, \BdToKSJpsimm, \BuToKstJpsiee, and \BuToKstJpsimm, the branching fractions of which are known to respect lepton universality to an excellent approximation \cite{PDG2020} and are taken to be equal for the muon and the electron decays of a given $B$ meson. The parameters \RinvKS and \RinvKstp are measured as double ratios
\begin{eqnarray}
\label{Eq:ratio}
    \RinvK &=& \frac{\mathcal{B}(\BToKee)}{\mathcal{B}(\BToKJpsiee)} \Big/ \frac{\mathcal{B}(\BToKmm)}{\mathcal{B}(\BToKJpsimm)}
    \nonumber \\ 
    &=& \left( \frac{N_{\text{sig}}^{ee}}{\epsilon_{\text{sig}}^{ee}} \cdot \frac{\epsilon_{\text{con}}^{ee}}{N_{\text{con}}^{ee}} \right)
    \Big/ \left( \frac{N_{\text{sig}}^{\mu\mu}}{\epsilon_{\text{sig}}^{\mu\mu}} \cdot \frac{\epsilon_{\text{con}}^{\mu\mu}}{N_{\text{con}}^{\mu\mu}} \right),
\end{eqnarray}
where $K^{(*)}$ is either a \KS or \Kstarp meson, $N$ is the measured yield, and $\epsilon$ is the total efficiency for signal (sig) and control (con) decays. The inverse ratio \RinvK is measured as its uncertainty better represents a Gaussian distribution due to the low yield of the electron decay mode. Many sources of systematic bias cancel in the ratio between the signal and control modes. The differential branching fractions of the signal electron modes are measured as
\begin{equation}
    \frac{\mathrm{d}\mathcal{B}\left(\BToKee \right)}{\mathrm{d}\qsq} = \frac{N_{\text{sig}}^{ee}}{\epsilon_{\text{sig}}^{ee}} \frac{\epsilon_{\text{con}}^{ee}}{N_{\text{con}}^{ee}} \frac{\mathcal{B}\left(\BToKJpsiee\right)}{q^2_{\rm \max} - q^2_{\rm \min}}.
\end{equation}

The LHCb detector is a single-arm forward spectrometer covering the pseudorapidity range $2 < \eta < 5$, 
described in detail in Refs.~\cite{LHCb-DP-2008-001,LHCb-DP-2014-002}. The simulated events used in this analysis are produced with the software described in Refs.~\cite{Sjostrand:2007gs,*Sjostrand:2006za,LHCb-PROC-2010-056,Lange:2001uf,Allison:2006ve,*Agostinelli:2002hh,LHCb-PROC-2011-006}. In particular, final-state radiation is simulated using \photos~\cite{davidson2015photos}.

The candidates used in the analysis must first pass a hardware trigger, which requires the presence of at least one muon with high transverse momentum, \pt, in the case of \BToKmm\ candidates, or in the case of \BToKee\ candidates at least one electron or hadron with large energy deposits in the electromagnetic calorimeter (ECAL) or hadronic calorimeter (HCAL), respectively. Further \BToKee\ candidates are selected where the hardware trigger requirements are satisfied by objects from the underlying $pp$ collision that do not form part of the reconstructed candidate. Candidates are then required to pass a software trigger, the first stage of which requires the presence of at least one track with high \pt that is well separated from the primary $pp$ interaction vertex (PV), followed by a second stage that imposes topological requirements on the final-state tracks to determine whether they are consistent with the decay of a $b$ hadron.

Muons are initially identified from tracks that penetrate the calorimeters and the iron absorber plates of the muon system and further separated from hadrons (primarily pions and kaons) by a multivariate classifier that combines information from the other subdetectors. Electrons are identified from tracks with an associated deposit of energy in the ECAL, and separated from hadrons using a similar multivariate classifier. 

Due to their small mass, electrons lose energy via bremsstrahlung radiation as they traverse the detector material, leading to a degradation in their energy and momentum resolution. A bremsstrahlung recovery procedure is used to identify energy deposits in the ECAL that are consistent with photons radiated from electron candidate tracks upstream of the magnet. This is done by extrapolating the direction of the electron track before the magnet to a position in the ECAL and then searching for energy deposits without associated tracks at that location. When such a deposit is identified, its energy is used to correct the electron's energy and momentum. This leads to an improvement in the $B$ candidate invariant mass resolution, although the resolution for electronic modes remains larger than for the equivalent muonic channels.

Candidate \KS mesons are reconstructed from two oppositely charged tracks identified as pions, using either a pair of tracks that originate in the vertex locator (long tracks), or two tracks that originate downstream of the vertex locator in the first silicon-strip detector (downstream tracks). Around a third of reconstructed \KS mesons are formed from long tracks. Candidate \BdToKSll\ decays are formed from two oppositely charged tracks identified as either muons or electrons combined with a candidate \KS meson. In the case of \BuToKstll\ candidates, the additional charged pion is required to result in a $\KS\pip$ mass within 300\mevcc of the \Kstarp mass \cite{PDG2020}. An estimated S-wave contribution in this \Kstarp mass window of approximately 22\% based on previous studies of \BdToKpimumu\ decays by the LHCb collaboration \cite{LHCb-PAPER-2016-012} is included in the analysis. When measuring the \BuToKstee\ differential branching fraction, the \BuToKstJpsiee\ control mode is selected with a $\KS\pip$ mass in the range $792-992$ \mevcc in order to be consistent with the selection used in previous measurements of $\mathcal{B}\left(\BuToKstJpsi\right)$ \cite{BaBar:2004htr, Belle:2002otd}, the world average of which is taken as external input \cite{PDG2020}.  Background is further suppressed by requirements on the quality of the $B$ decay vertex, the flight distance significance of the $B$ candidate, how consistent the $B$ candidate is with having originated at the PV, the invariant masses of the \KS, \Kstarp and \B candidates, and the \pt and separation from the PV of the final-state tracks. In an additional step, the invariant masses of $B$ candidates are recalculated with the \KS meson mass constrained to its measured value \cite{PDG2020}, leading to an improvement in the $B$ mass resolution.

Various requirements on decay kinematics, decay time, and particle identification (PID) information are used to reject potential background originating from misidentified $b$-hadron ($H_b$) decays. Background to \BdToKSll\ decays includes \mbox{$H_b \rightarrow h h' \ellp \ellm$} decays (where $h$ refers to a hadron), \LbToLll\ and \BdToXDKSY\ decays, where both $X$ and $Y$ represent either an electron-neutrino pair or a pion. Background to \BuToKstll\ decays includes \BdToKSll\ decays combined with a random additional pion, \BuToKstJpsill\ and \BuToKstpsitwosll\ decays where the companion pion from the \Kstarp is swapped with a lepton from the \jpsi or \psitwos meson, \mbox{$H_b \rightarrow h h' \pi \ellp \ellm$} decays, decays with $\Lz$ baryons in the final state, and \BuToXDKSY\ decays. These selection requirements reduce all these background sources to levels that have a negligible (sub-percent) effect on the measured signal yields. The \BdToKSpipi\ and \BuToKstpipi\ decays, where the pions are misidentified as electrons, are significantly reduced by the electron particle identification requirements and included as components in the mass fits.

Multivariate classifiers based on boosted decision tree (BDT) algorithms \cite{Breiman} are used to suppress background from coincidental track combinations (combinatorial background). Separate classifiers are trained for each signal decay mode, the two data-taking periods (Run 1, Run 2) and whether the \KS meson was reconstructed from long or downstream tracks. Each classifier is trained on a signal sample of simulated \BToKll\ decays, and a background sample taken from data with a reconstructed invariant $B$ mass greater than 5500\mevcc and \qsq regions consistent with either a \jpsi or \psitwos meson removed. The classifiers combine information on the $B$ candidate's fit quality, distance of closest approach to the PV, flight distance, \pt and decay time; the decay time of the dilepton pair; the \pt and decay time of the \KS candidate; how isolated the $B$, dilepton and \KS candidates are from other tracks in the event; and the distance of closest approach to the PV of long tracks forming the \KS candidate. Requirements on the classifier outputs are optimised to provide the maximum signal significance, defined as $S/\sqrt{S+B}$, where $S$ is the expected signal calculated from the control mode yield in data, the signal-to-control mode efficiency ratio from simulation and the signal-to-control mode branching fraction ratio \cite{PDG2020}, and $B$ is the background yield in the signal region, extrapolated from a fit to the data mass sidebands.

\begin{figure}[t]
    \centering
     \includegraphics[width=0.48\linewidth]{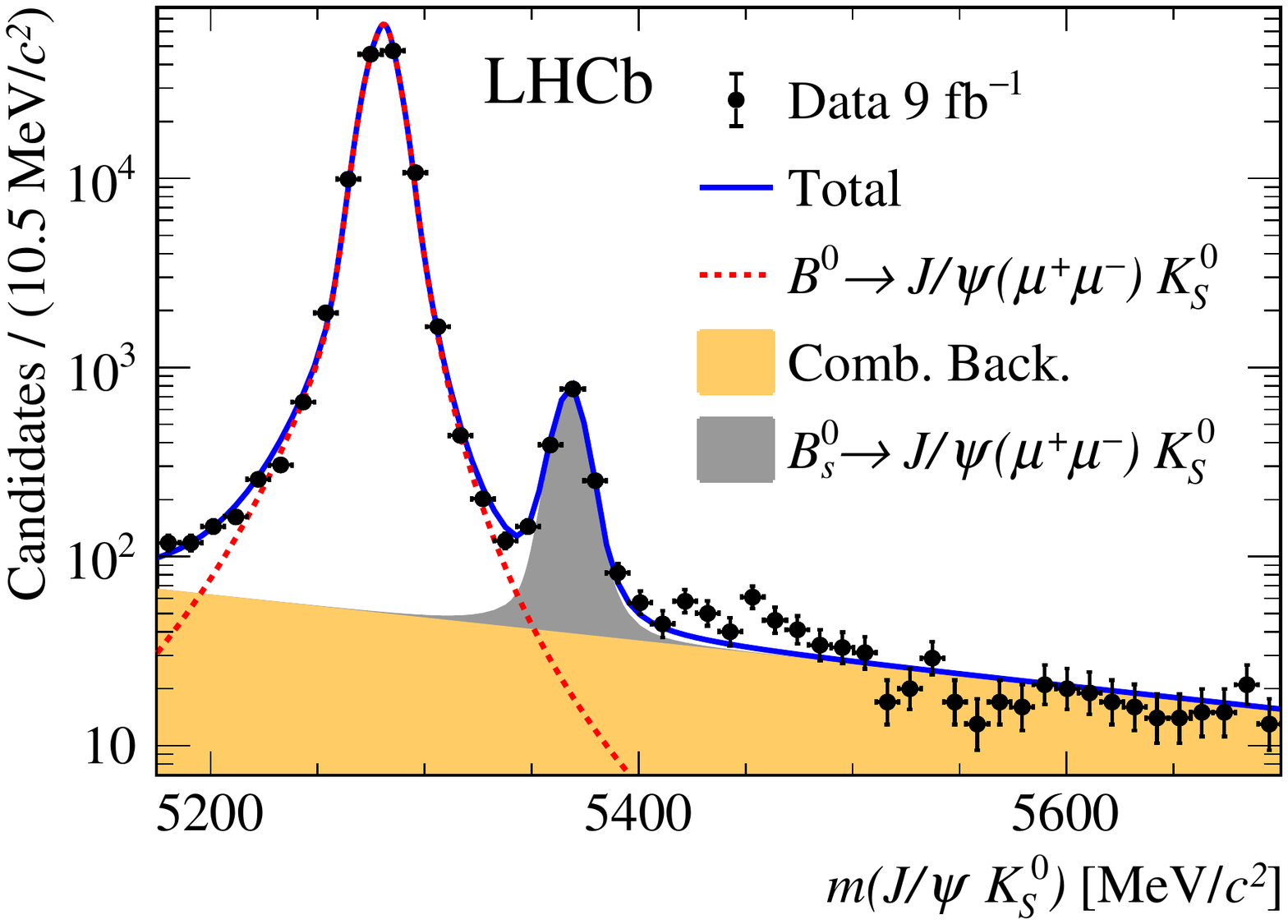}\hfill
     \includegraphics[width=0.48\linewidth]{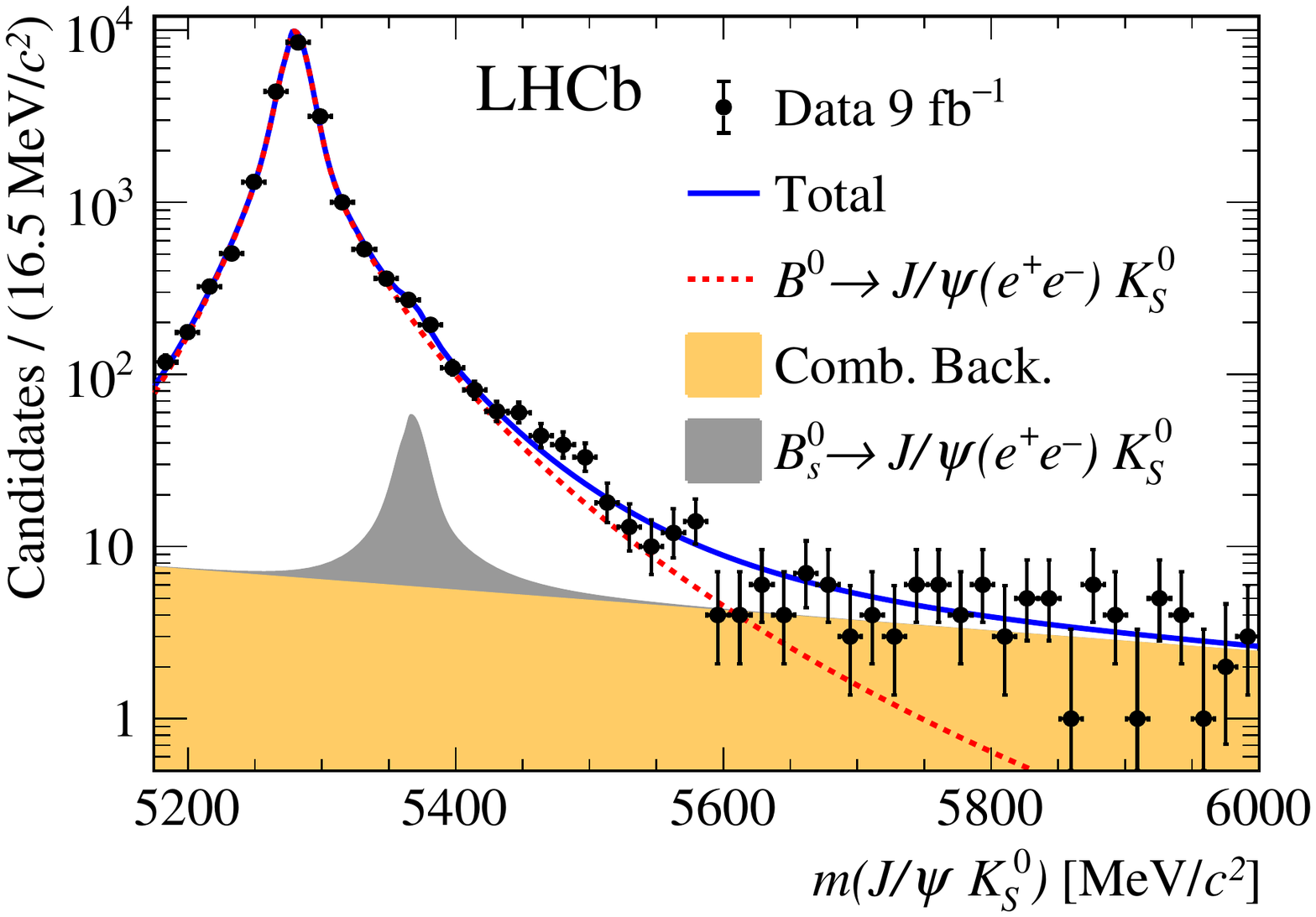}
     
     \includegraphics[width=0.48\linewidth]{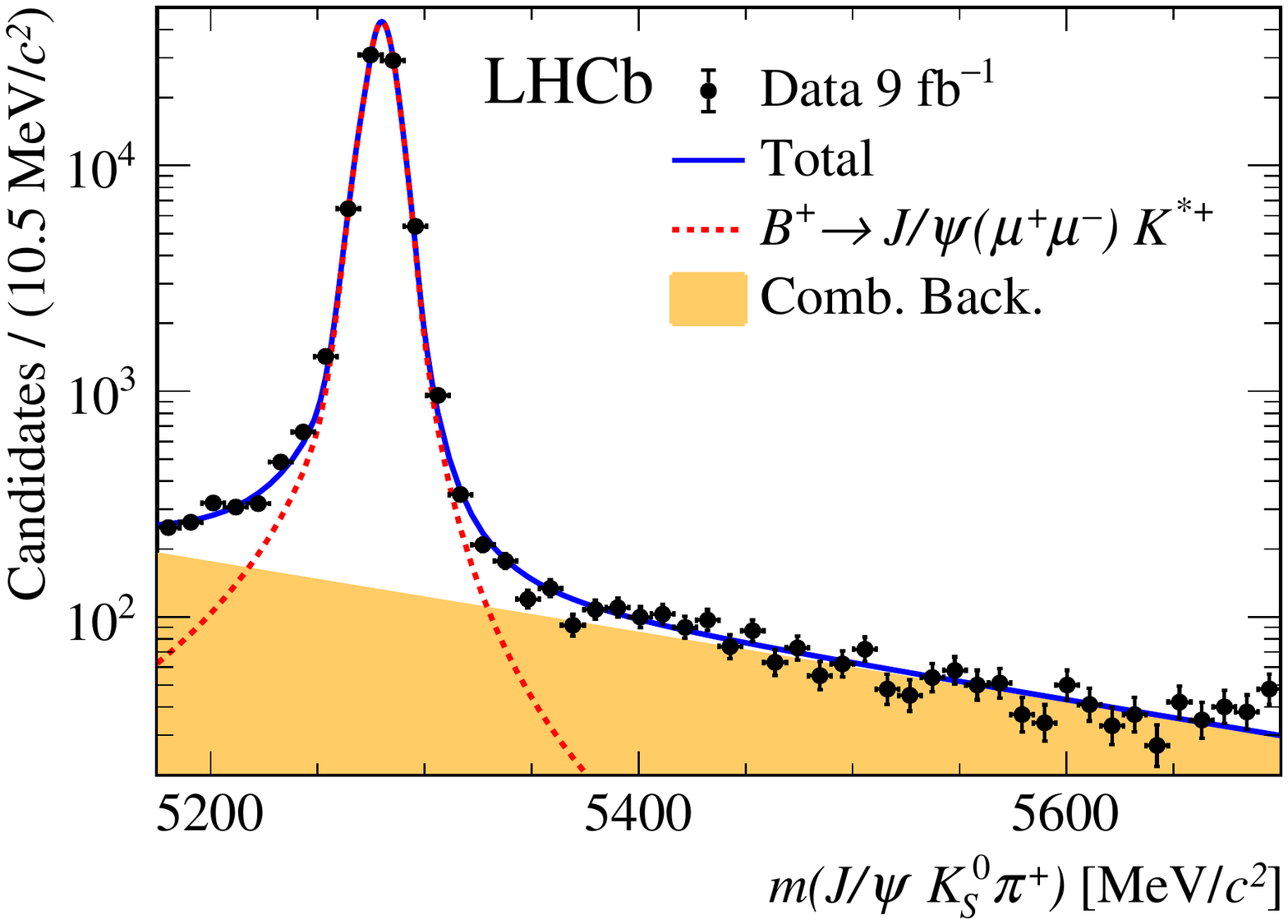}\hfill
     \includegraphics[width=0.48\linewidth]{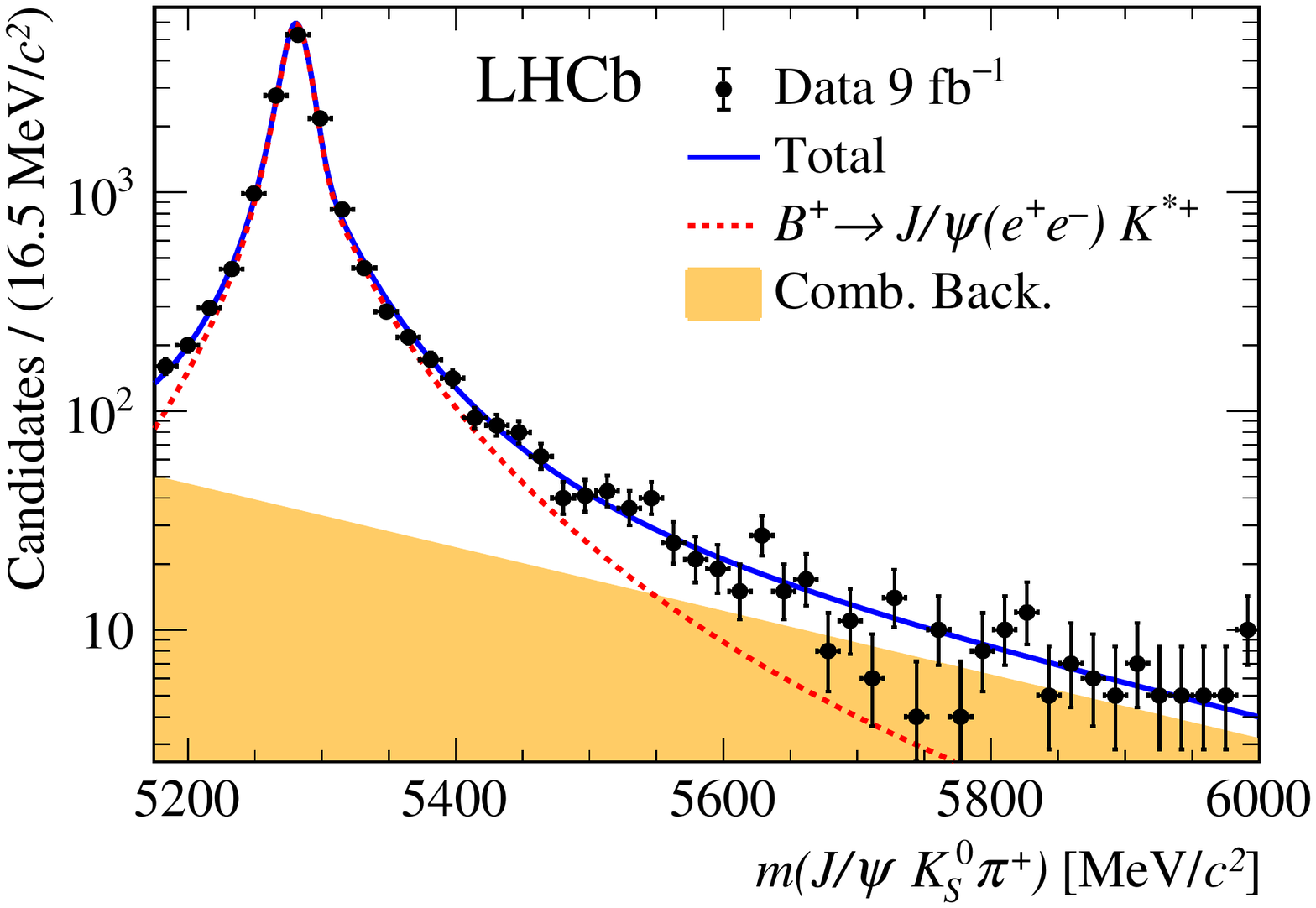}
\caption{\small Distributions of (top left)~$\jpsi(\mup\mun)\KS$ mass, (top right)~$\jpsi(\ep\en)\KS$ mass, (bottom left)~$\jpsi(\mup\mun)\KS\pip$ mass and (bottom right)~$\jpsi(\ep\en)\KS\pip$ mass with the fit models used to determine the control mode yields.}\label{fig:controlfit}
\end{figure}

The muon and electron control modes are selected in the ranges of \mbox{$8.98 < \qsq < 10.21$\gevgevcccc} and \mbox{$6.0 < \qsq < 11.0$\gevgevcccc}, respectively, and the spectra of their invariant masses $m(\jpsi \KS)$ and $m(\jpsi \KS \pip)$ are shown in Fig. \ref{fig:controlfit}. Their yields are determined using fits to the $B$ candidate mass, calculated with the masses of the \KS and \jpsi mesons constrained to their measured values \cite{PDG2020}, improving the resolution. The mass distribution of each control mode is modelled with a sum of two Crystal Ball functions \cite{Skwarnicki:1986xj} with common mean and opposite-side power law tails (referred to as a double Crystal Ball or DCB), with parameters fixed to values obtained from fits to simulated events. The muon control modes are modelled using a single DCB function, while the electron control modes are modelled using a sum of three DCB functions, where the shape of each component and their relative fractions are determined from fits to simulation in three different categories according to the number of bremsstrahlung photons added to the candidate's electrons: 0, 1, or $\geq 2$. In the fit to data, shifts in the DCB means and widths are allowed to vary freely with respect to those obtained from simulation in order to accommodate data-simulation differences in the mass scale and resolution. Combinatorial background is modelled with an exponential function. In the case of the \Bz control modes, \BsToKSJpsill\ decays are modelled with the \Bz DCB function with its mean offset by the measured $m_{\Bs}-m_{\Bz}$ mass difference \cite{PDG2020} and with its yield allowed to vary freely. The lower limit of the mass range excludes partially reconstructed background from higher $K^*$ resonances, which are not modelled. The structures visible in the range $5400 \mevcc < m(\jpsi\KS) < 5500 \mevcc$ are due to small amounts of residual contamination from \Lb decays, which have a negligible effect on the measured yields of \BdToKSJpsimm\ and \BdToKSJpsiee\ decays. The results of the fit are shown in Fig.~\ref{fig:controlfit} and the yields of \BdToKSJpsimm, \BdToKSJpsiee, \BuToKstJpsimm, and \BuToKstJpsiee\ decays are found to be \mbox{$118\,750 \pm 360$}, \mbox{$21\,080 \pm 170$}, \mbox{$75\,420 \pm 290$}, and \mbox{$14\,330 \pm 170$}, respectively.

\begin{figure}[t]
    \centering
     \includegraphics[width=0.48\linewidth]{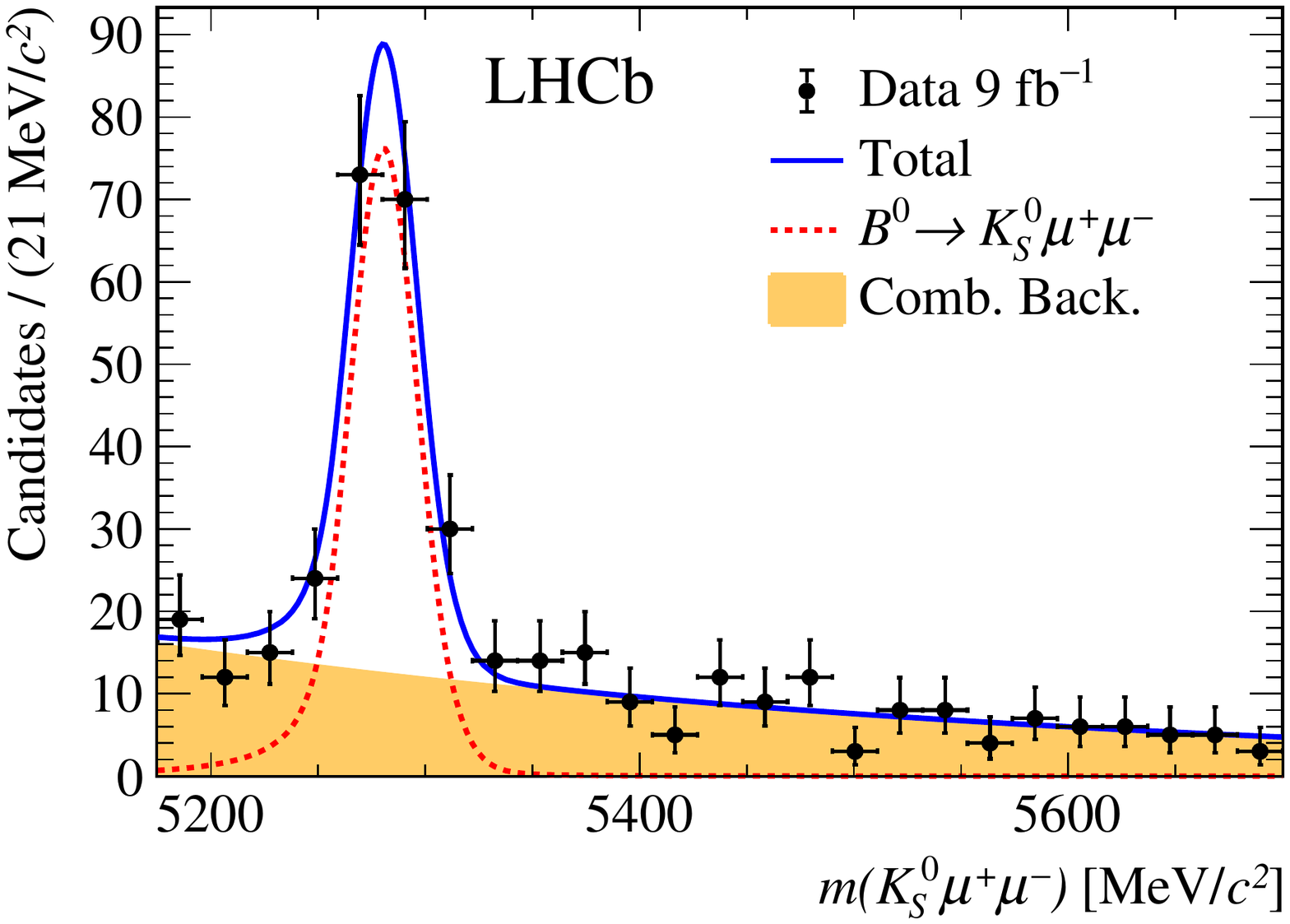}\hfill
     \includegraphics[width=0.48\linewidth]{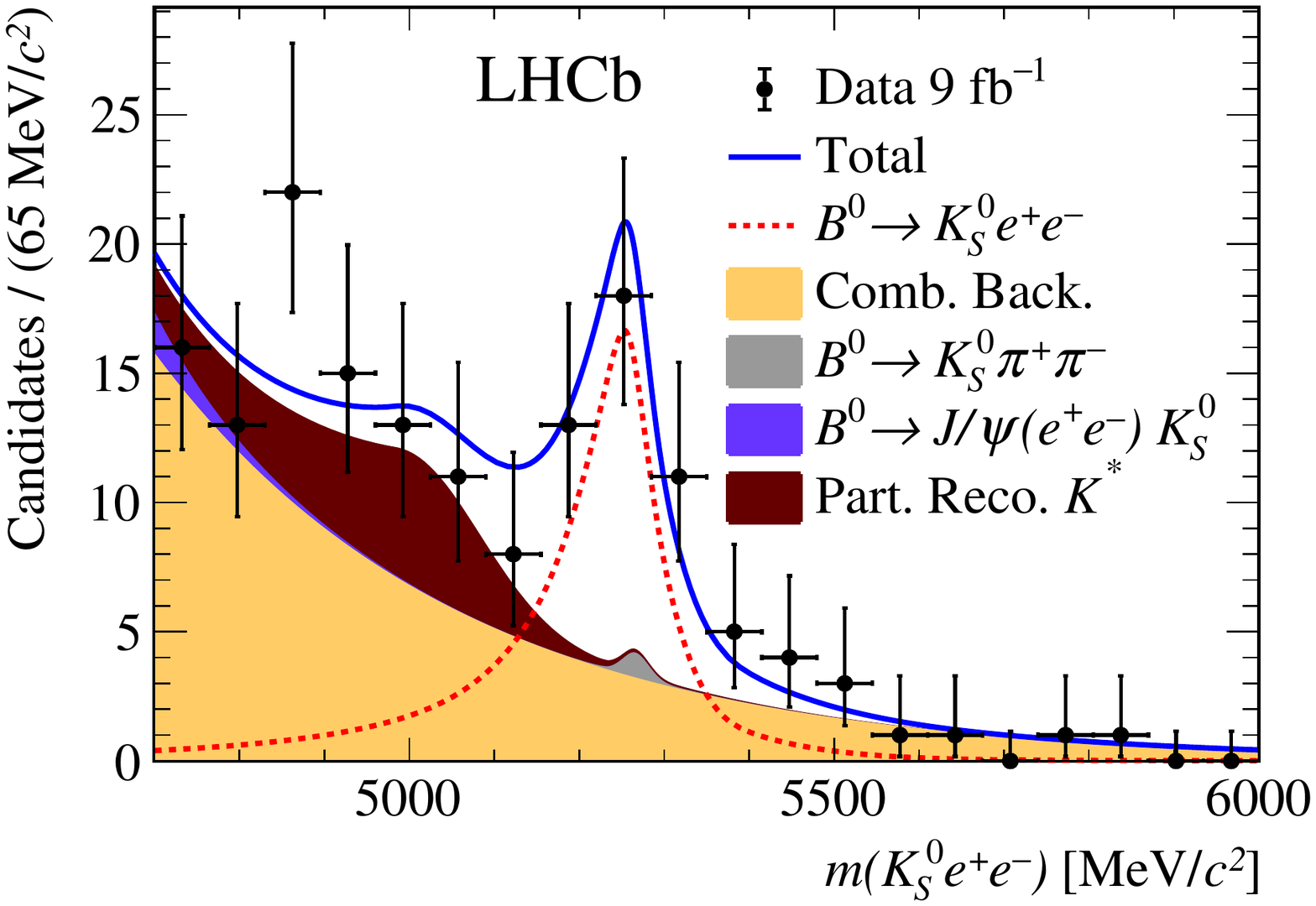}
     \includegraphics[width=0.48\linewidth]{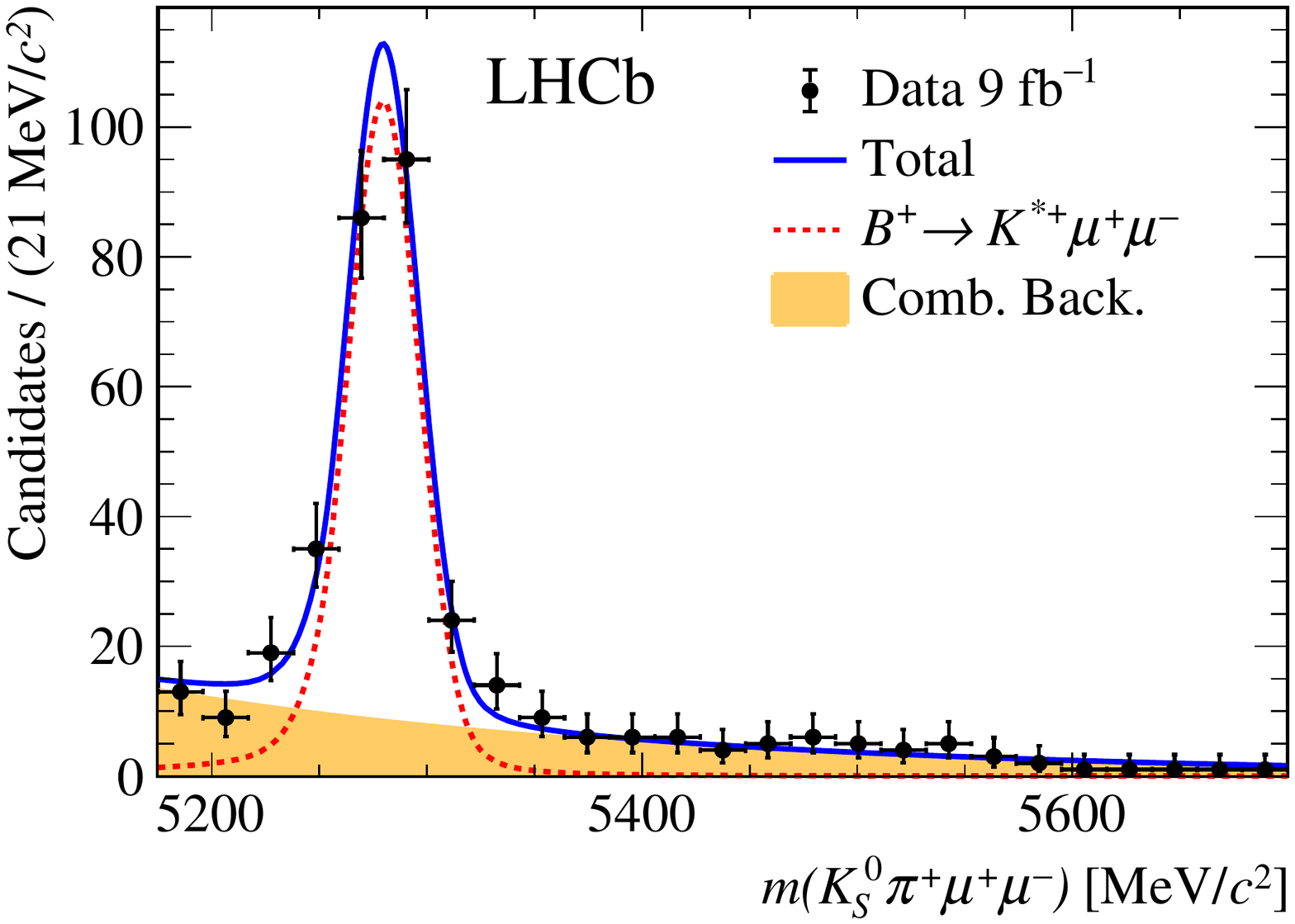}\hfill
     \includegraphics[width=0.48\linewidth]{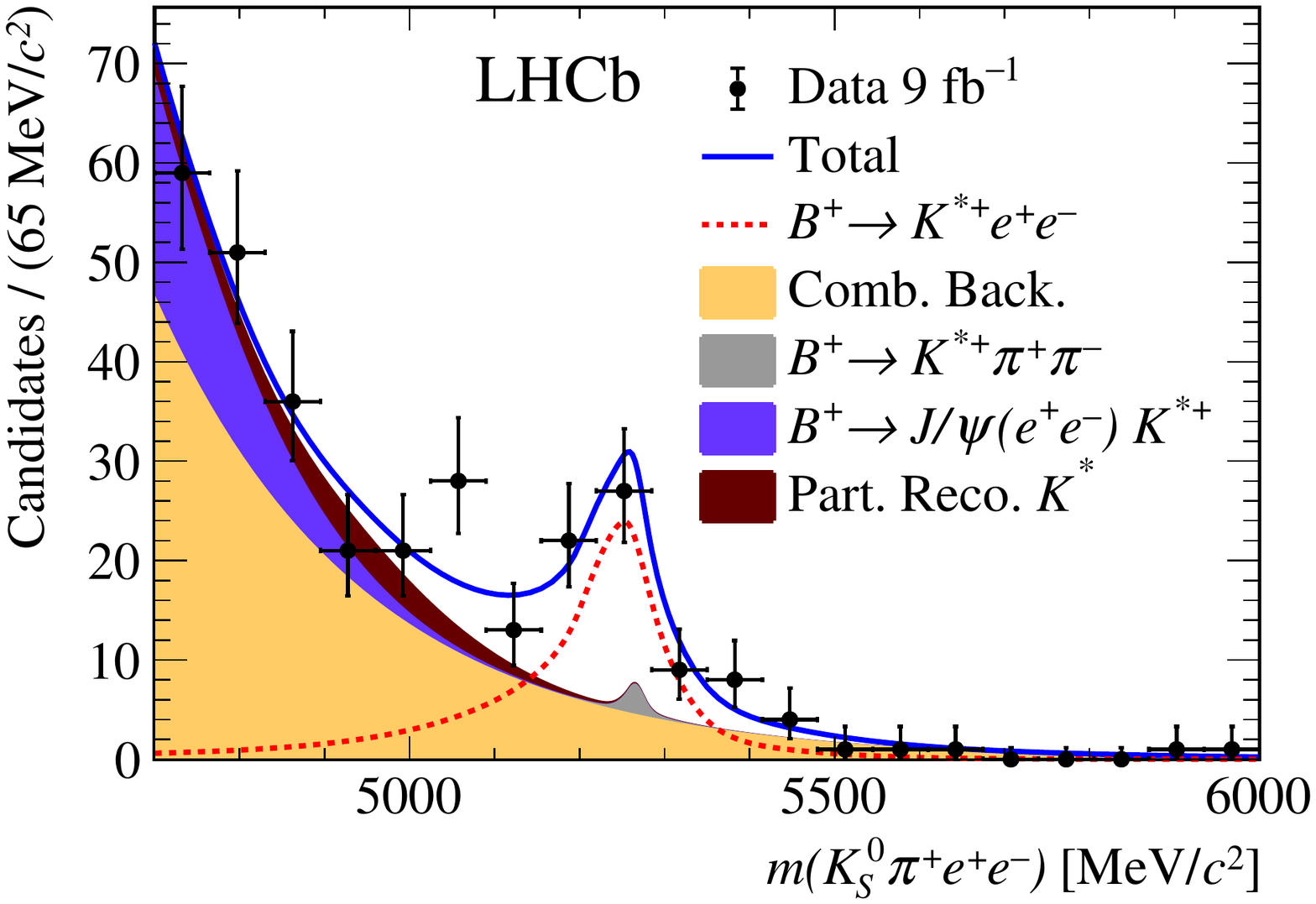}
\caption{\small Distributions of (top left) $\KS\mup\mun$ and (top right) $\KS\epem$ mass with the fit models used to determine the \BdToKSmm\ yield and \RinvKS, and (bottom left) $\KS\pip\mup\mun$ and (bottom right) $\KS\pip\epem$ mass with the fit models used to determine the \BuToKstmm\ yield and \RinvKstp.}\label{fig:signalmassfits}
\end{figure}

The spectra of the invariant masses $m(\KS\mup\mun)$, $m(\KS\ep\en)$, $m(\KS\pip\mup\mun)$, and $m(\KS\pip\ep\en)$ of the muon and electron signal modes are shown in Fig. \ref{fig:signalmassfits}. The yields of \BdToKSmm\ and \BuToKstmm\ decays are determined using fits to the $\KS\mup\mun$ and $\KS\pip\mup\mun$ mass distributions. The \BdToKSmm\ and \BuToKstmm\ signal decays are modelled using DCB functions where the shape parameters are determined from fits to simulation, with shifts in their means and widths taken from the corresponding control mode fits to data. Combinatorial background is modelled using an exponential function, while partially reconstructed background is excluded by the lower mass limit.

The ratios and branching fractions are determined using fits to $\KS\epem$ and $\KS\pip\epem$ mass spectra, with the \KS candidate's mass constrained to its measured value. The \BdToKSee\ and \BuToKstee\ signal decays are modelled using the sum of three DCB functions, each corresponding to different numbers of recovered bremsstrahlung photons. The DCB parameters are taken from simulation with shifts in the means and widths taken from the control mode fits to data without a \jpsi mass constraint. Partially reconstructed background from higher $K^{*}$ resonances is modelled using a DCB function with shape parameters constrained from simulation in the \BdToKSee\ fit, and Gaussian kernel density estimations (KDE) determined from simulation in the \BuToKstee\ fit, with their yields allowed to vary freely. Leakage from the \jpsi control modes into the signal region is modelled using KDE functions, with their yields constrained based on the control mode fits and the efficiency in simulation. Residual contamination from \BdToKSpipi\ and \BuToKstpipi\ decays is modelled in each fit by a DCB function determined from simulated events, with its yield constrained using the control mode yields, the control mode and background branching fractions, and the efficiencies taken from simulation. Combinatorial background is modelled with an exponential function.

The efficiencies used in the measurements of the ratios and differential branching fractions are calculated using simulation, to which various corrections are applied to improve the agreement with data. The PID efficiencies for each channel are calculated from calibration data samples of electrons, muons and pions, and are applied as per-candidate weights to the simulation. Similarly, the electron tracking efficiency is corrected using calibration samples. The \pt and pseudorapidity of the $B$ mesons generated by \pythia 8 \cite{Sjostrand:2007gs}, and the occupancy of the underlying events are corrected by comparing their distributions between data and simulation using the muon control modes to calculate per-candidate weights, which are applied to both electron and muon samples. Similarly, the trigger efficiency is corrected by comparing the efficiency as a function of the \pt of the muons, the transverse energy of the electrons and pions, and the \pt of the $B$ meson, between control mode data and simulation. Further weights are applied to correct any residual mismodelling of the BDT classifier response. Finally, the simulated \qsq distribution is corrected using control mode data to account for the larger observed resolution in data.

Multiple sources of systematic uncertainty are evaluated, the largest of which comes from the statistical uncertainties of the efficiencies due to the sizes of the available samples of simulated events. These affect the \RinvK ratios and the differential branching fractions at 2--3\%. The next largest are associated with the mass fit models, in particular the limited size of the simulation samples used to determine the shape parameters, and the choices of models used for the partially reconstructed and \jpsi leakage background, which affect the observables by 1--2\%. The remaining sources of systematic uncertainty are found to be close to or below the 1\% level. These include: the limited amount of data and simulation used to calculate correction weights; the choices of binning schemes used to evaluate the PID efficiency weights and potential biases in this procedure due to correlations in the PID response between the two electrons; the choice of methods used to calculate the trigger efficiency; imperfect modelling of the muon track reconstruction efficiency; residual mismodelling of the BDT classifier response in simulation; residual bias in the fitting procedure evaluated using pseudoexperiments, and residual contamination from \BdToXDKSY\ and \BuToXDKSY\ decays. Other sources of residual background
contamination were found to have a negligible effect on the
measurements. All systematic uncertainties, as well as the statistical precision of the efficiencies and various parameters used in the maximum likelihood fits, are included as Gaussian constraints in the fit. Finally, the statistical precision on each observable is scaled by factors determined from pseudoexperiments (with values in the range 1.01--1.02), in order to guarantee proper coverage.

A number of checks are performed to ensure that efficiencies are accurately estimated. The ratios \RinvpsitwosK, where the signal modes in Eq.~(\ref{Eq:ratio}) are substituted for \BToKpsitwosmm\ and \BToKpsitwosee\ decays, selected in the \qsq ranges $[12.86,14.33]$\gevgevcccc and $[11.0, 15.0]$\gevgevcccc, respectively, are found to be \mbox{$\RinvpsitwosKS = 1.014 \pm 0.030 \left(\text{stat.}\right) \pm 0.020 \left(\text{syst.}\right)$} and \mbox{$\RinvpsitwosKstp = 1.017 \pm 0.045 \left(\text{stat.}\right) \pm 0.023 \left(\text{syst.}\right)$}, consistent with unity as expected due to lepton universality in \jpsi and \psitwos decays \cite{LHCb-PAPER-2021-004}. Additionally, the single ratios
\begin{equation}
    \rinvJpsiK \equiv \frac{ \mathcal{B}\left(\BToKJpsiee\right) } {\mathcal{B}\left(\BToKJpsimm\right)} = \frac{N_{\text{con}}^{ee}}{N_{\text{con}}^{\mu\mu}} \frac{\epsilon_{\text{con}}^{\mu\mu}}{\epsilon_{\text{con}}^{ee}},
\end{equation}
which do not benefit from cancellation of systematic biases between signal and control modes, and are therefore a stringent check of the efficiencies, are found to be \mbox{$\rinvJpsiKS = 0.977 \pm 0.008 \left(\text{stat.}\right) \pm 0.027 \left(\text{syst.}\right)$} and \mbox{$\rinvJpsiKstp = 0.965 \pm 0.011 \left(\text{stat.}\right) \pm 0.032 \left(\text{syst.}\right)$}, again consistent with unity in both cases. Furthermore, differential measurements of \rJpsiK as functions of a range of variables that are differently distributed in the signal and control decays are performed. The most powerful of these tests measures \rJpsiK as a function of the response of a BDT classifier trained on simulated candidates to distinguish signal and control decays. All these differential distributions are found to be flat within the statistical precision, providing further confidence that the yields and efficiencies are well estimated \cite{Supplemental}.

In order to avoid experimenter’s bias, the results of the analysis and the electron signal mode mass spectra were not examined until the full procedure had been finalised and checks had been performed to ensure that the muon signal mode branching fractions were in agreement between the different data-taking years and also with the results of the previous measurements made by LHCb \cite{LHCb-PAPER-2014-006}. The fits to the \BdToKSee, \BdToKSmm, \BuToKstee\ and \BuToKstmm\ invariant-mass spectra are shown in Fig.~\ref{fig:signalmassfits}. The fitted yields of \BdToKSee\, \BdToKSmm\, \BuToKstee\ and \BuToKstmm\ decays are $45 \pm 10$, $155 \pm 15$, $67 \pm 13$ and $221 \pm 17$, respectively. The ratios \RinvKS and \RinvKstp are measured to be
\begin{align*}
    \RinvKS=\hspace{0.1cm}& 1.51\,^{+0.40}_{-0.35} \left(\text{stat.}\right) ^{+0.09}_{-0.04} \left(\text{syst.}\right), \\
    \RinvKstp=\hspace{0.1cm}& 1.44\, ^{+0.32}_{-0.29} \left(\text{stat.}\right) ^{+0.09}_{-0.06} \left(\text{syst.}\right),
\end{align*}
in the \qsq ranges $[1.1, 6.0]$ \gevgevcccc and $[0.045, 6.0]$ \gevgevcccc, respectively.
These ratios are consistent with the SM at $1.5$ and $1.4$ standard deviations, respectively, evaluated using Wilks' theorem \cite{Wilks:1938dza}. To aid comparison with other lepton-universality ratios, \RKS and \RKstp are calculated by inverting the results above, yielding
\begin{align*}
    \RKS=\hspace{0.1cm}& 0.66\,^{+0.20}_{-0.14} \left(\text{stat.}\right) ^{+0.02}_{-0.04} \left(\text{syst.}\right), \\ 
    \RKstp=\hspace{0.1cm}& 0.70\,^{+0.18}_{-0.13} \left(\text{stat.}\right) ^{+0.03}_{-0.04} \left(\text{syst.}\right). 
\end{align*}
The differential branching fractions of the signal electron decays are found to be
\begin{align*}
     \frac{\mathrm{d}\mathcal{B}\left(\BdToKzee \right)}{\mathrm{d}\qsq}=&\, \left(2.6 \pm 0.6 \left(\text{stat.}\right) \pm 0.1  \left(\text{syst.}\right) \right) \times 10^{-8} \hspace{0.1cm} \gev^{-2}c^4, \\
    \frac{\mathrm{d}\mathcal{B}\left(\BuToKstee \right)}{\mathrm{d}\qsq}=&\, \left(9.2\,^{+1.9}_{-1.8} \left(\text{stat.}\right) ^{+0.8}_{-0.6}  \left(\text{syst.}\right)\right)  \times 10^{-8} \hspace{0.1cm} \gev^{-2}c^4,
\end{align*}
in the \qsq ranges $[1.1, 6.0]$ \gevgevcccc and $[0.045, 6.0]$ \gevgevcccc and where the significances of the \BdToKSee\ and \BuToKstee\ decays evaluated using Wilks' theorem \cite{Wilks:1938dza} are 5.3$\sigma$ and 6.0$\sigma$, respectively. Since the control mode branching fraction of \BdToKzJpsi\ decays $(8.91 \pm 0.21) \times 10^{-4}$ \cite{PDG2020} is used, the differential branching fraction of \BdToKzee\ instead of \BdToKSee\ decays is reported. A combination of the \RinvKS and \RinvKstp measurements is performed using the {\sc flavio} software package~\cite{Straub:2018kue} to fit for a single muon-specific Wilson coefficient $C_9^{\text{NP}} = -C_{10}^{\text{NP}}$, while fixing all other Wilson coefficients to their SM values. This scenario is used in several existing fits to \bTosll data, and is chosen specifically as the ratios \RKbrackst have poor sensitivity in discriminating between the Wilson Coefficients $C_9$ and $C_{10}$. The fit results in $C_9^{\text{NP}} = -C_{10}^{\text{NP}}=-0.8^{+0.4}_{-0.3}$ and a significance of 2.0 standard deviations with respect to the SM under this specific scenario. It should be noted that this fit is model-dependent and the result could change if data-driven estimates of the hadronic uncertainties were used.

These measurements constitute the most precise tests of lepton universality in \BdToKSll\ and \BuToKstll\ decays to date, the most precise measurements of their differential branching fractions at low \qsq, and the first observations of \BdToKSee\ and \BuToKstee\ decays. While these measurements are individually consistent with the SM, the central values exhibit the same deficit of muonic decays relative to electronic decays as seen in the other lepton universality tests performed by the LHCb collaboration \cite{LHCb-PAPER-2017-013, LHCb-PAPER-2021-004, LHCb-PAPER-2019-040}.

% Do not include this in any draft (just for information in the template)
% Comment this in for paper drafts; do not include this in analysis note, conference and figure reports
\section*{Acknowledgements}
%
% These Acknowledgements valid from 3-May-2019
%
\noindent We express our gratitude to our colleagues in the CERN
accelerator departments for the excellent performance of the LHC. We
thank the technical and administrative staff at the LHCb
institutes.
We acknowledge support from CERN and from the national agencies:
CAPES, CNPq, FAPERJ and FINEP (Brazil); 
MOST and NSFC (China); 
CNRS/IN2P3 (France); 
BMBF, DFG and MPG (Germany); 
INFN (Italy); 
NWO (Netherlands); 
MNiSW and NCN (Poland); 
MEN/IFA (Romania); 
MSHE (Russia); 
MICINN (Spain); 
SNSF and SER (Switzerland); 
NASU (Ukraine); 
STFC (United Kingdom); 
DOE NP and NSF (USA).
We acknowledge the computing resources that are provided by CERN, IN2P3
(France), KIT and DESY (Germany), INFN (Italy), SURF (Netherlands),
PIC (Spain), GridPP (United Kingdom), RRCKI and Yandex
LLC (Russia), CSCS (Switzerland), IFIN-HH (Romania), CBPF (Brazil),
PL-GRID (Poland) and NERSC (USA).
We are indebted to the communities behind the multiple open-source
software packages on which we depend.
Individual groups or members have received support from
ARC and ARDC (Australia);
AvH Foundation (Germany);
EPLANET, Marie Sk\l{}odowska-Curie Actions and ERC (European Union);
A*MIDEX, ANR, IPhU and Labex P2IO, and R\'{e}gion Auvergne-Rh\^{o}ne-Alpes (France);
Key Research Program of Frontier Sciences of CAS, CAS PIFI, CAS CCEPP, 
Fundamental Research Funds for the Central Universities, 
and Sci. \& Tech. Program of Guangzhou (China);
%Key Research Program of Frontier Sciences of CAS, CAS PIFI,
%Thousand Talents Program, and Sci. \& Tech. Program of Guangzhou (China);
RFBR, RSF and Yandex LLC (Russia);
GVA, XuntaGal and GENCAT (Spain);
the Leverhulme Trust, the Royal Society
 and UKRI (United Kingdom).

\appendix
\clearpage

\section{Supplemental material}
\label{sec:Supplemental}
\begin{figure}[!htb]
    \centering
    \includegraphics[width=0.48\linewidth]{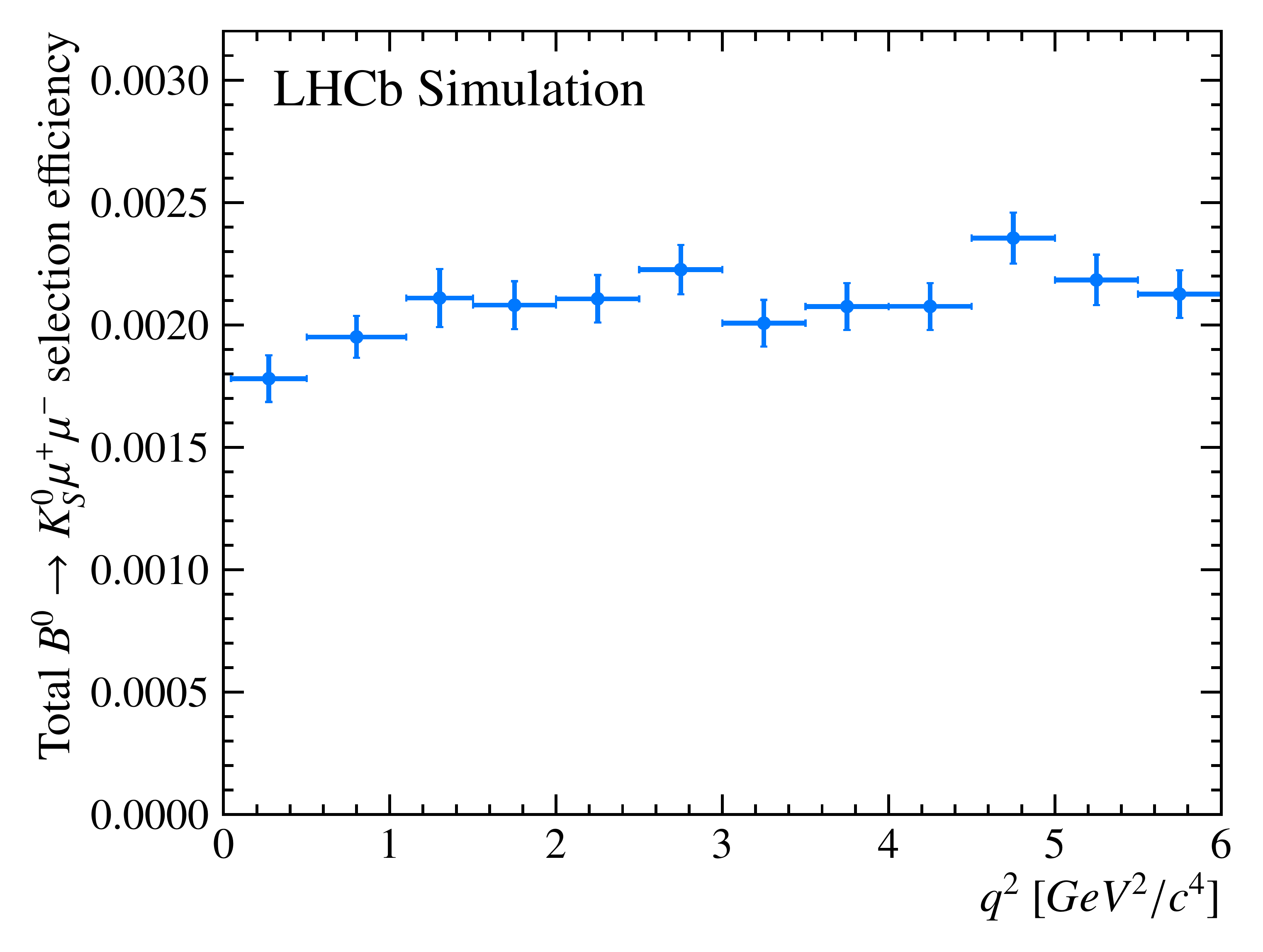}%
    \includegraphics[width=0.48\linewidth]{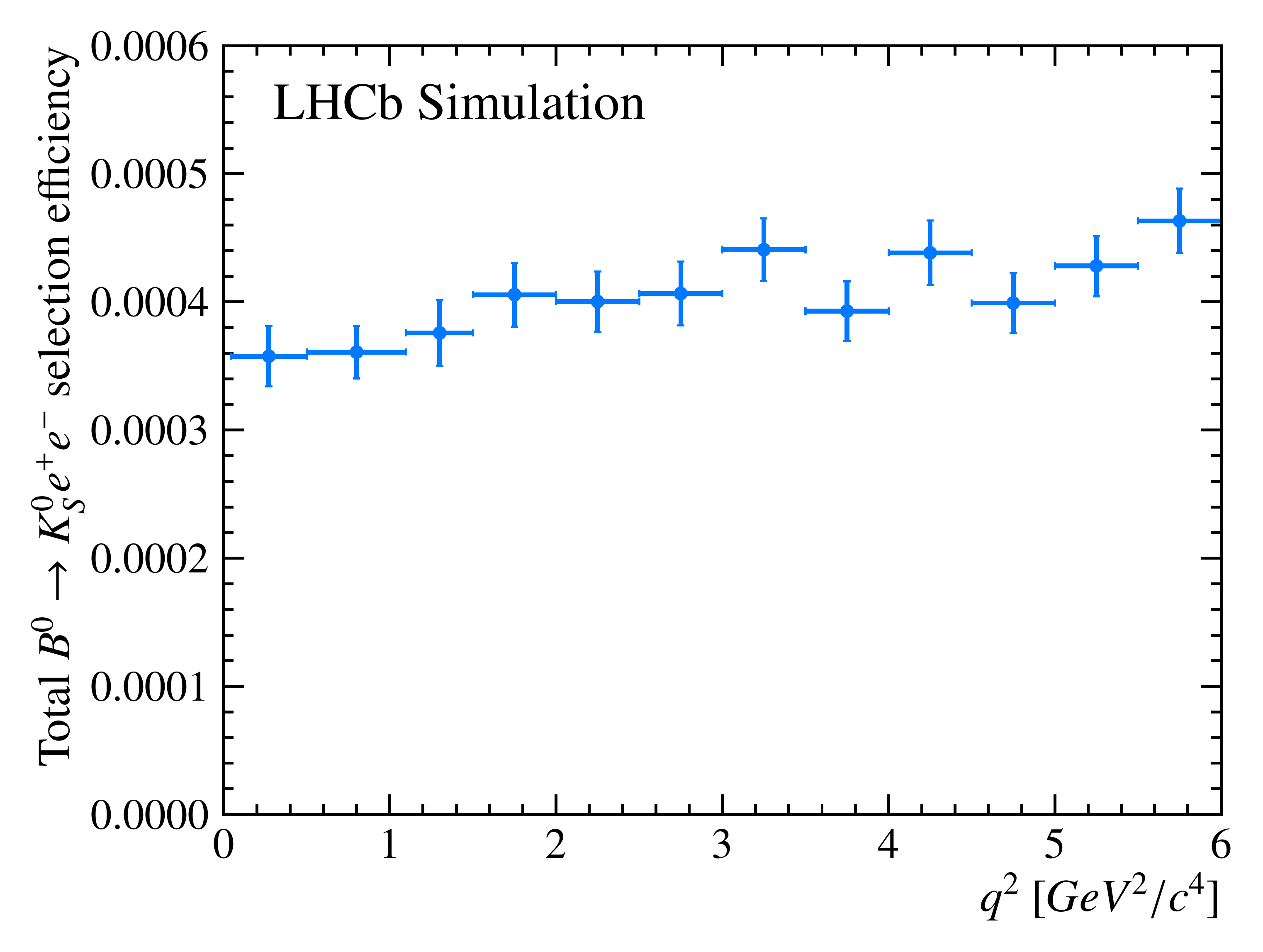}%

    \includegraphics[width=0.48\linewidth]{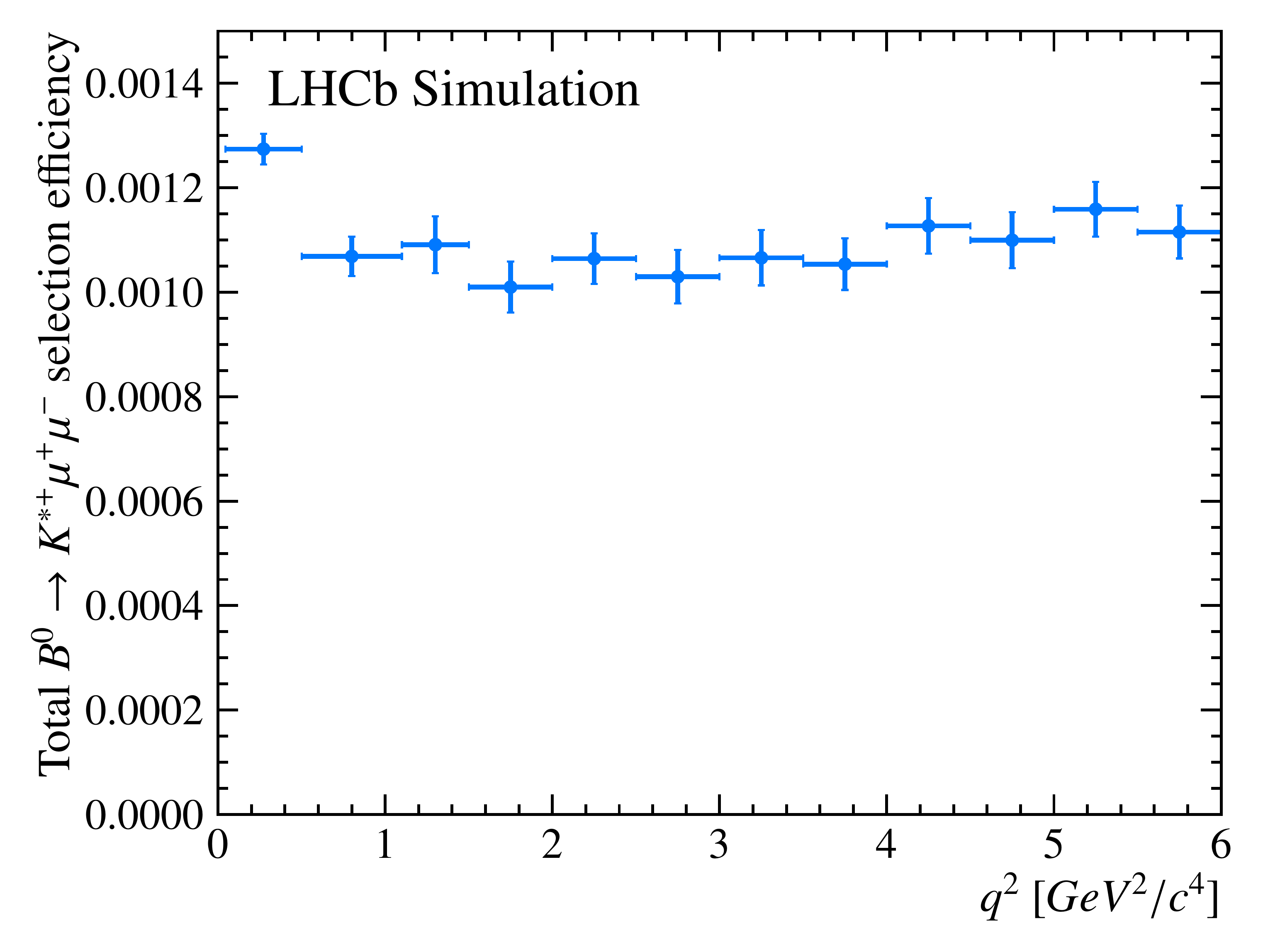}%
    \includegraphics[width=0.48\linewidth]{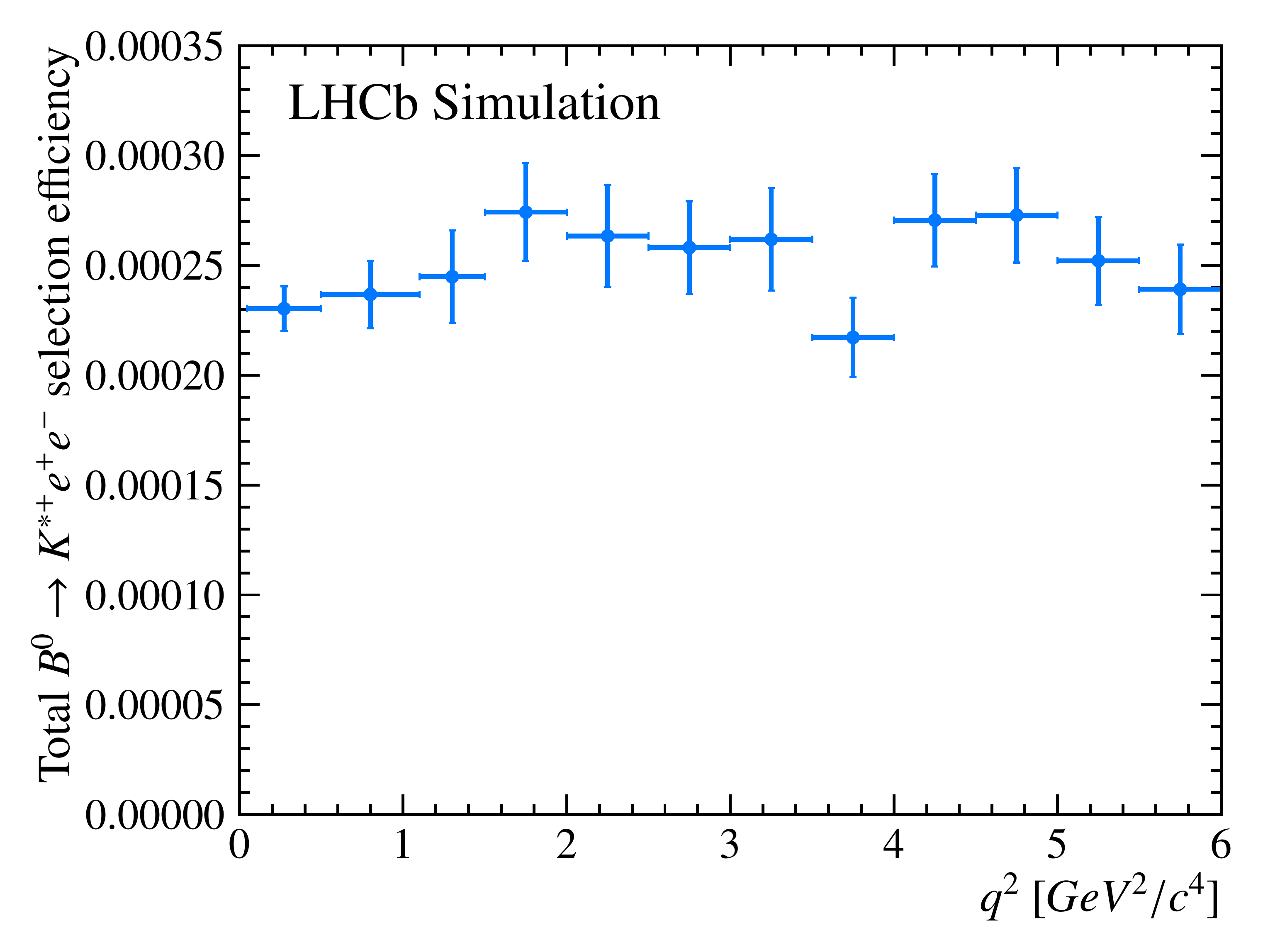}%
    
    \caption{
        \small %captions should be a little bit smaller than main text
        Total selection efficiencies as a function of \qsq, for (top left) \BdToKSmm, (top right) \BdToKSee, (bottom left) \BuToKstmm, and (bottom right) \BuToKstee\ decays. The efficiencies are evaluated using simulated candidates that have been corrected to improve agreement with data. These efficiencies account for detector resolution effects including the effect of bremsstrahlung, which may cause a candidate to be reconstructed in a different bin of \qsq than would be dictated by its `true' value.}
    \label{fig:supplementary_eff_profiles}
\end{figure}

\begin{figure}[!htb]
    \centering
    \includegraphics[width=0.48\linewidth]{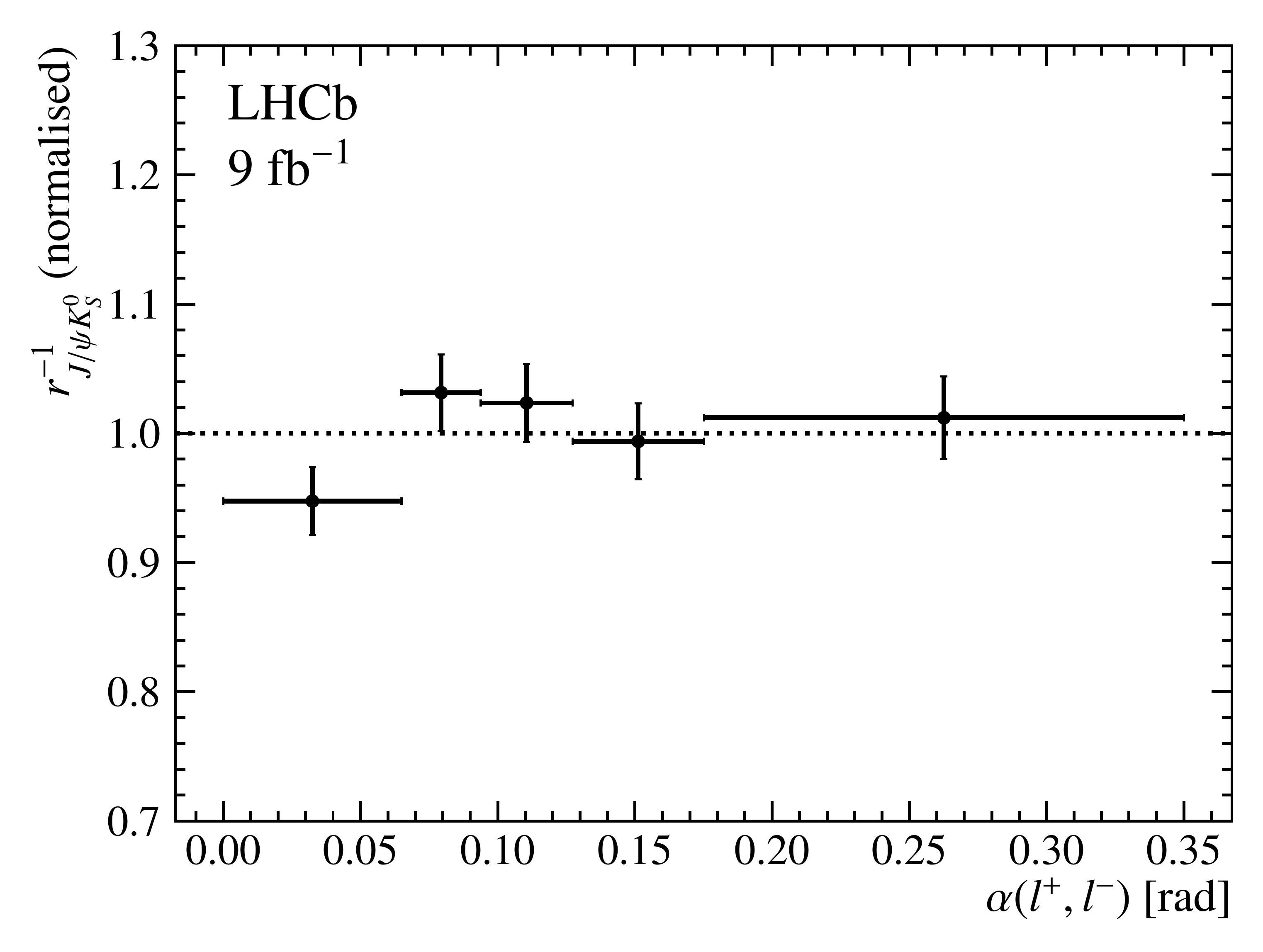}%
    \includegraphics[width=0.48\linewidth]{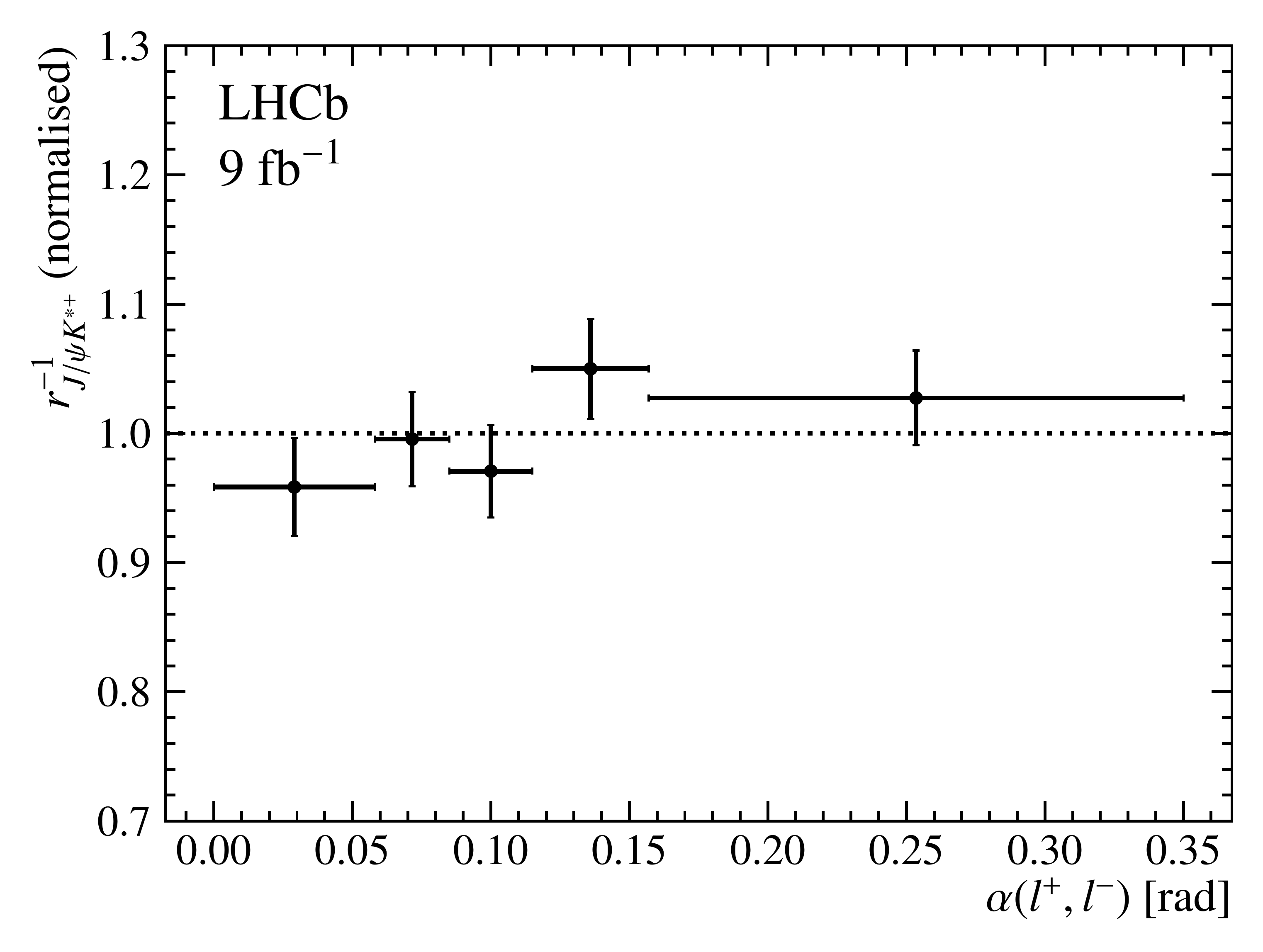}%
    
    \caption{
        \small
        Single ratios (left) \rinvJpsiKS and (right) \rinvJpsiKstp as a function of the opening angle of the two leptons, normalised to their average value.}
    \label{fig:supplementary_rjpsi_dilepton_opening_angle}
\end{figure}

\begin{figure}[!htb]
    \centering
    \includegraphics[width=0.48\linewidth]{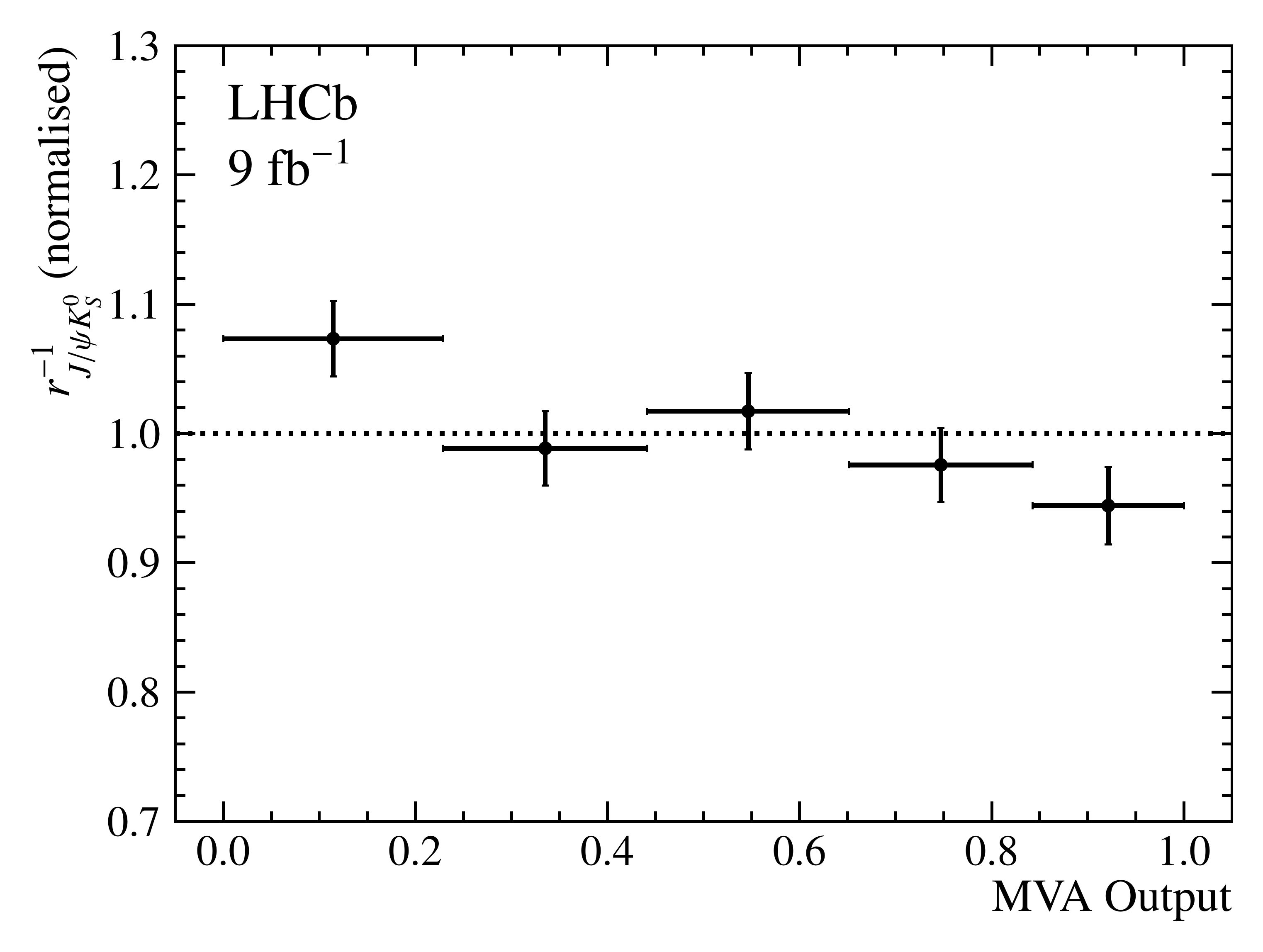}%
    \includegraphics[width=0.48\linewidth]{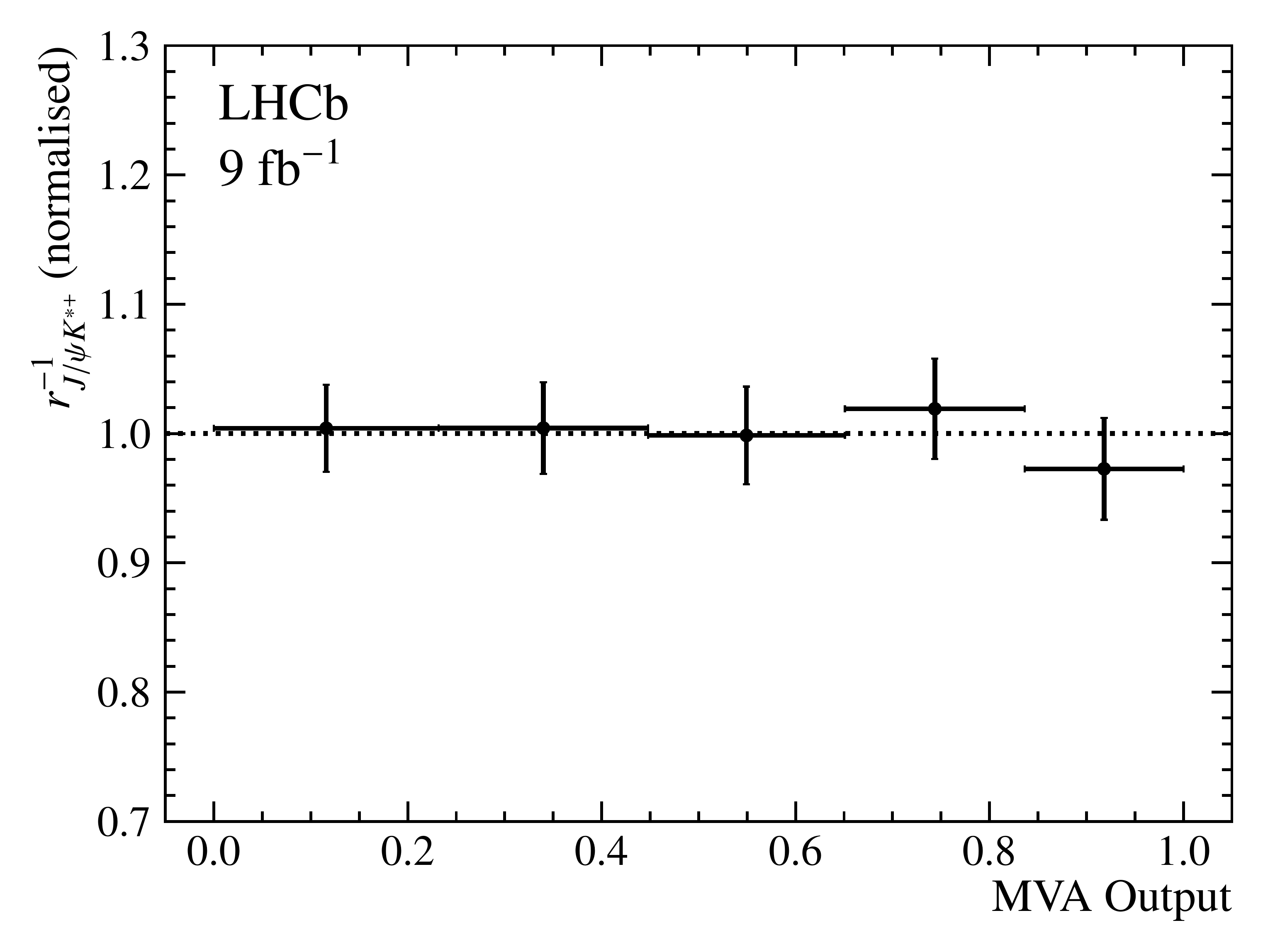}%
    
    \caption{
        \small
        Single ratios (left) \rinvJpsiKS and (right) \rinvJpsiKstp as a function of the output of an MVA trained to separate the signal and control modes, normalised to their average value.}
    \label{fig:supplementary_rjpsi_shell_clas}
\end{figure}

\begin{figure}[!htb]
    \centering
    \includegraphics[width=0.48\linewidth]{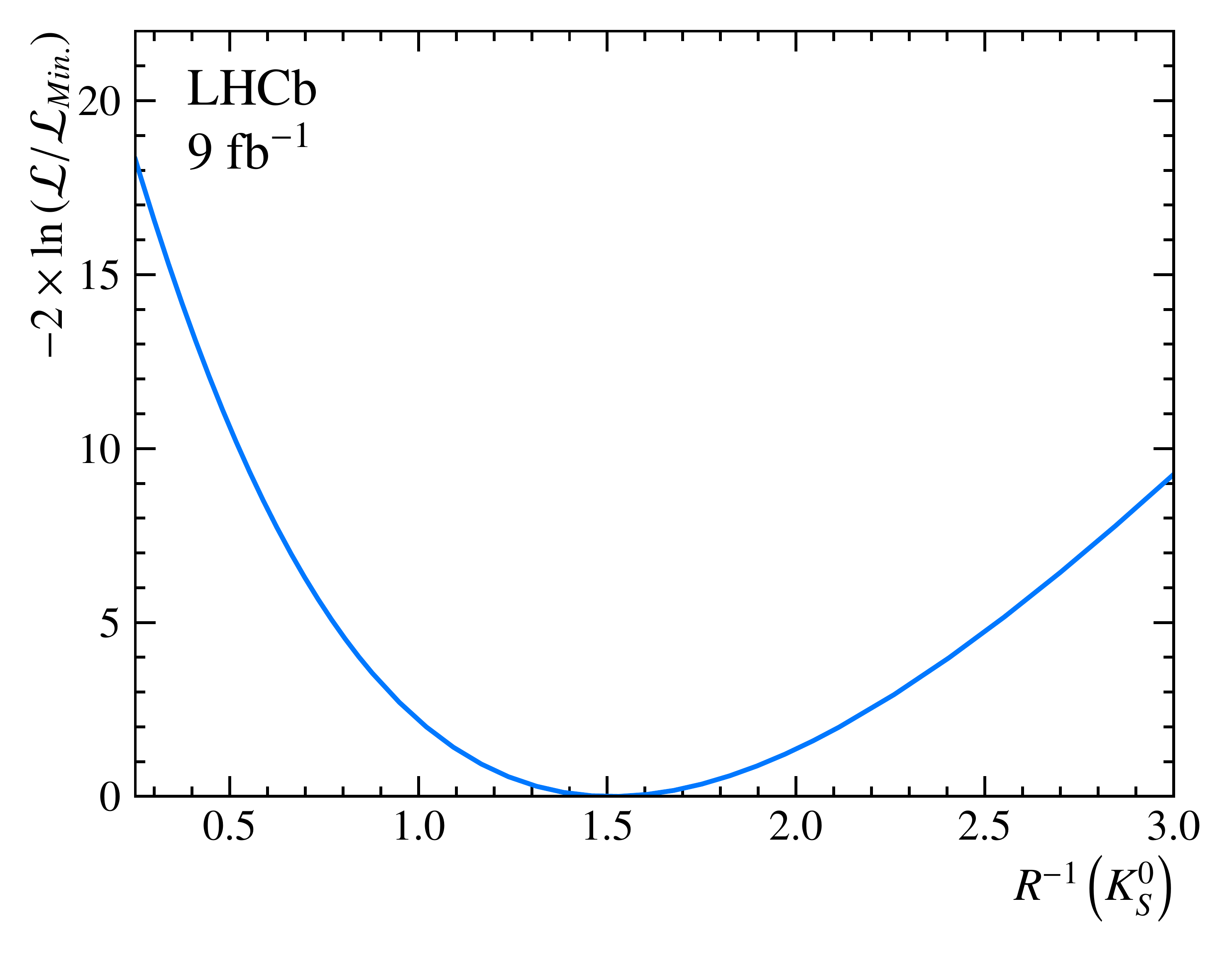}%
    \includegraphics[width=0.48\linewidth]{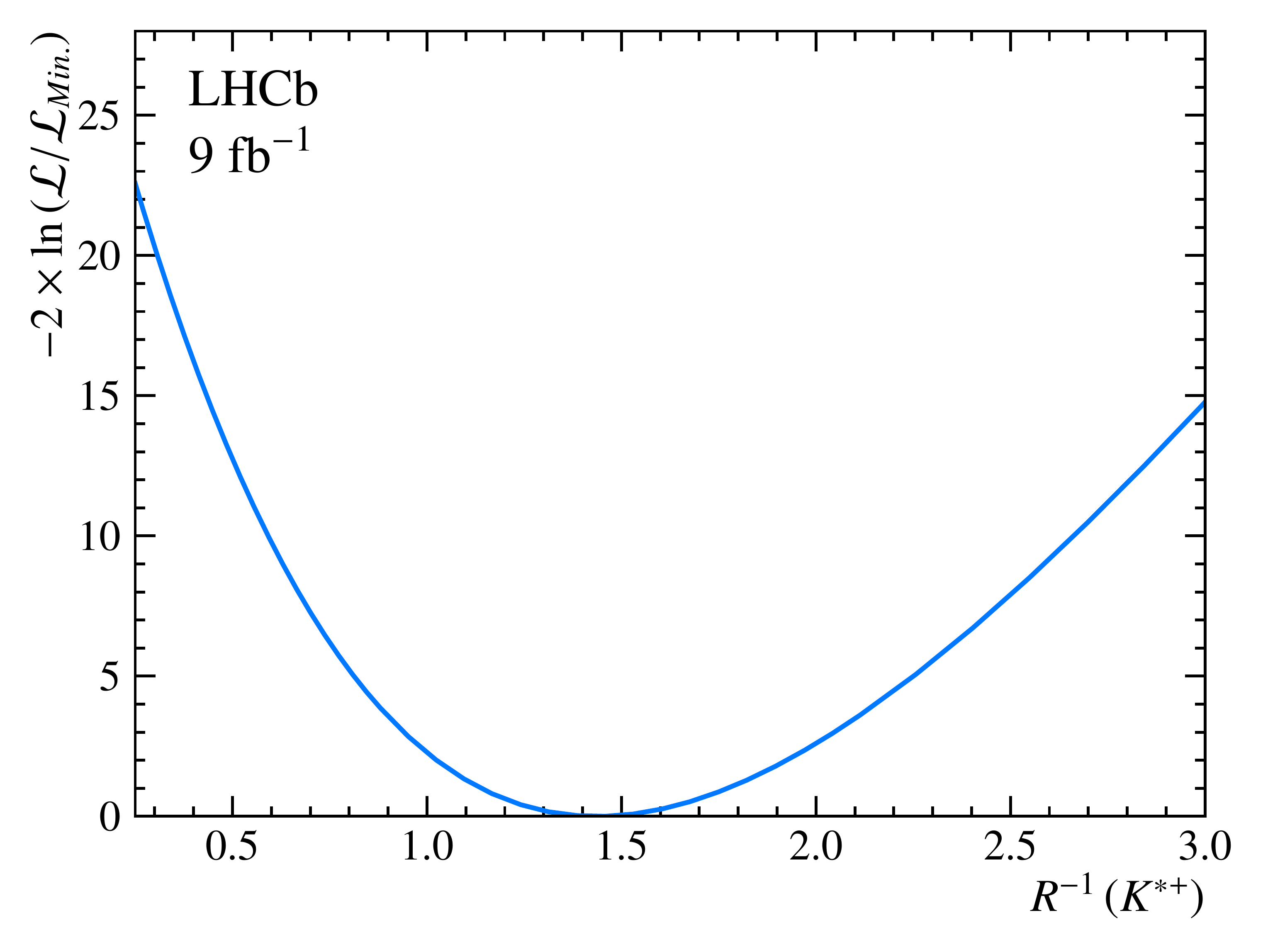}
    
    \caption{
        \small
        Scans of the profile likelihoods as a function of (left) \RinvKS, and (right) \RinvKstp.}
    \label{fig:supplementary_likelihood_scans_rx}
\end{figure}

\begin{figure}[!htb]
    \centering
    \includegraphics[width=0.48\linewidth]{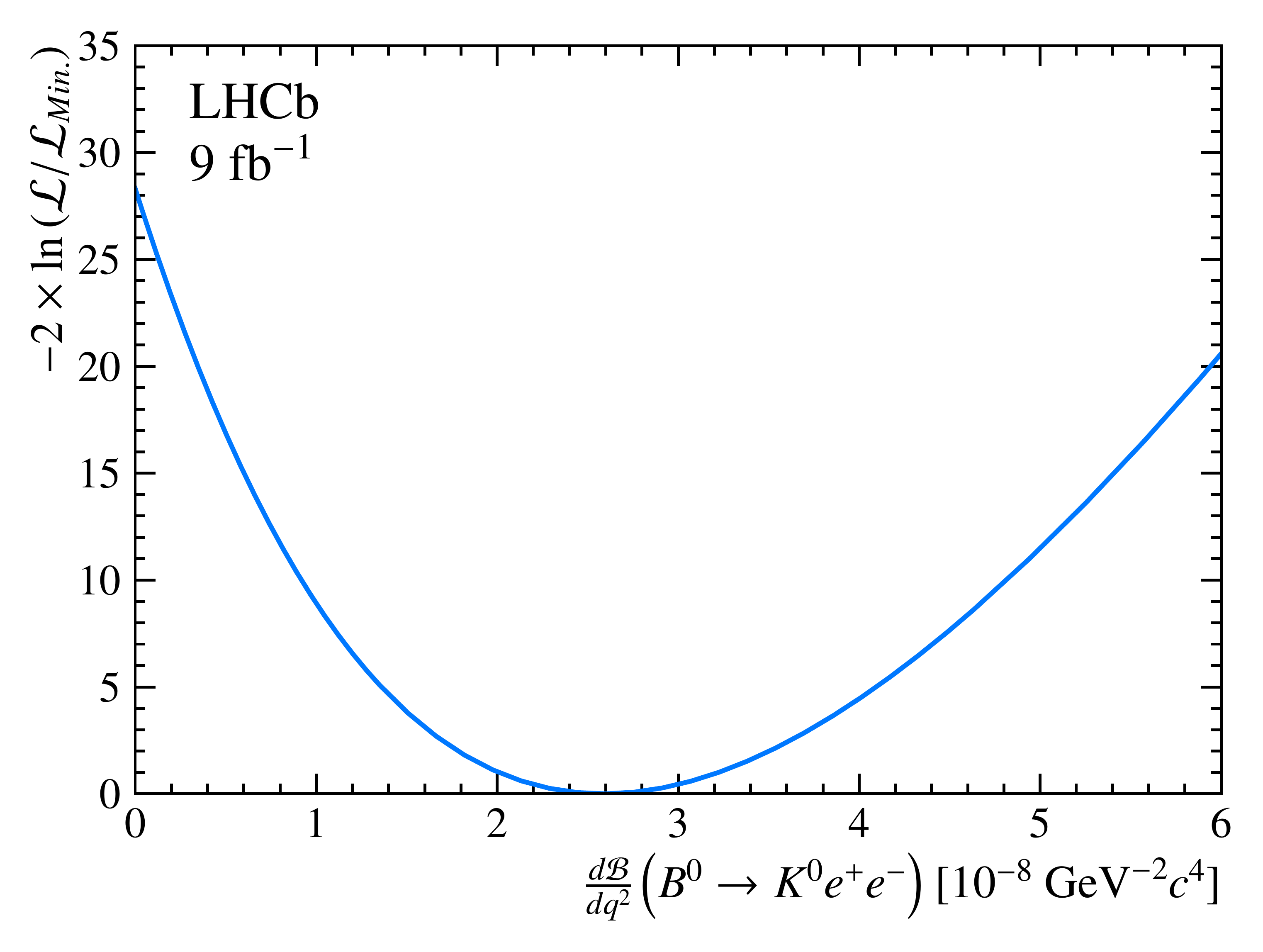}%
    \includegraphics[width=0.48\linewidth]{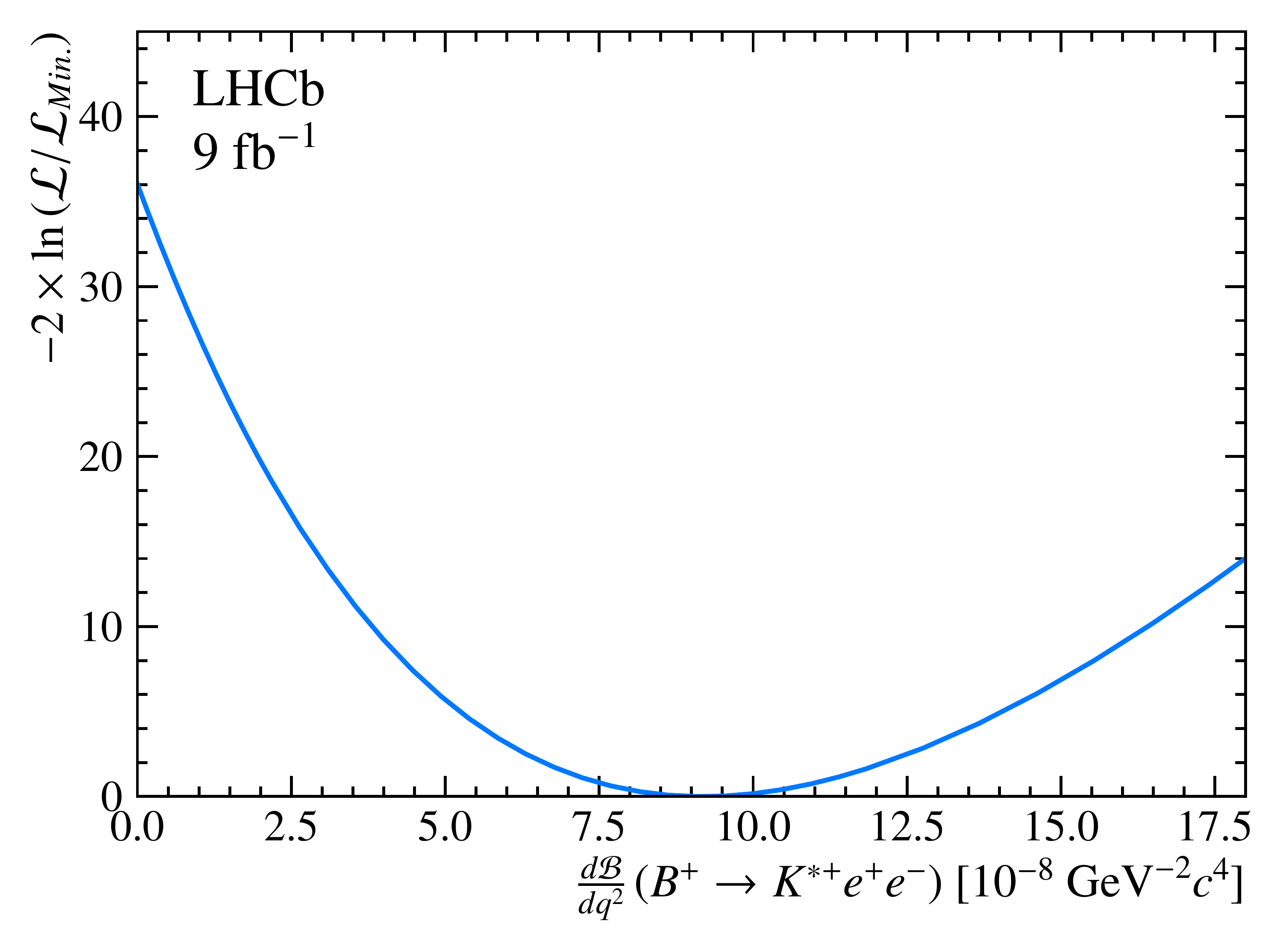}%
    
    \caption{
        \small
        Scans of the profile likelihoods as a function of the differential branching fractions for (left) \BdToKzee\ decays, and (right) \BuToKstee\ decays.}
    \label{fig:supplementary_likelihood_scans_rare_ee_bfs}
\end{figure}

\begin{figure}[!htb]
    \centering
    \includegraphics[width=0.75\linewidth]{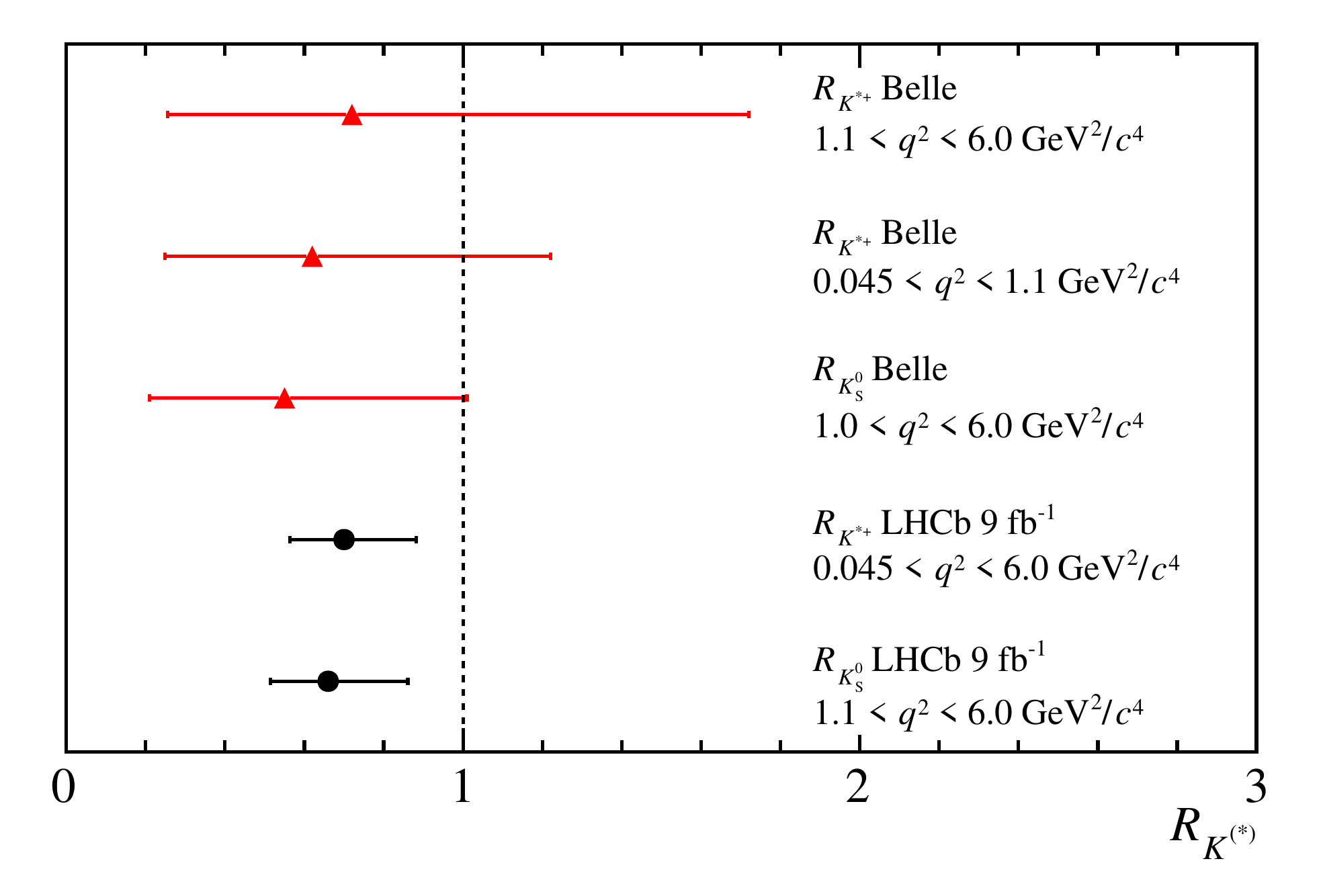}%
    \caption{
        \small
        Measurements of \RKS and \RKstp performed by the LHCb and Belle collaborations \cite{Belle:2019oag, BELLE:2019xld}.}
    \label{fig:supplementary_r_comparisons}
\end{figure}

\begin{figure}[t]
    \centering
     \includegraphics[width=0.48\linewidth]{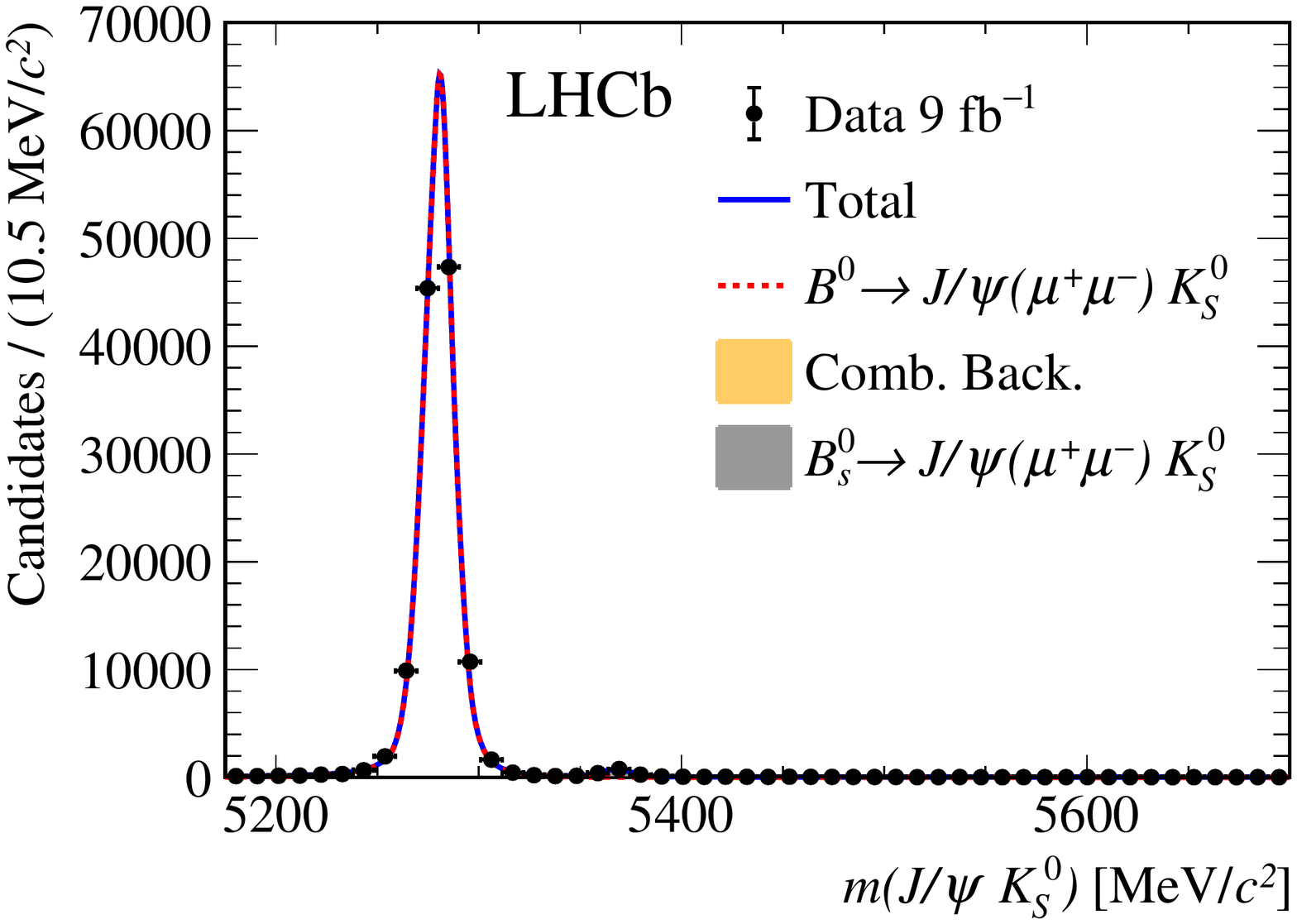}
     \includegraphics[width=0.48\linewidth]{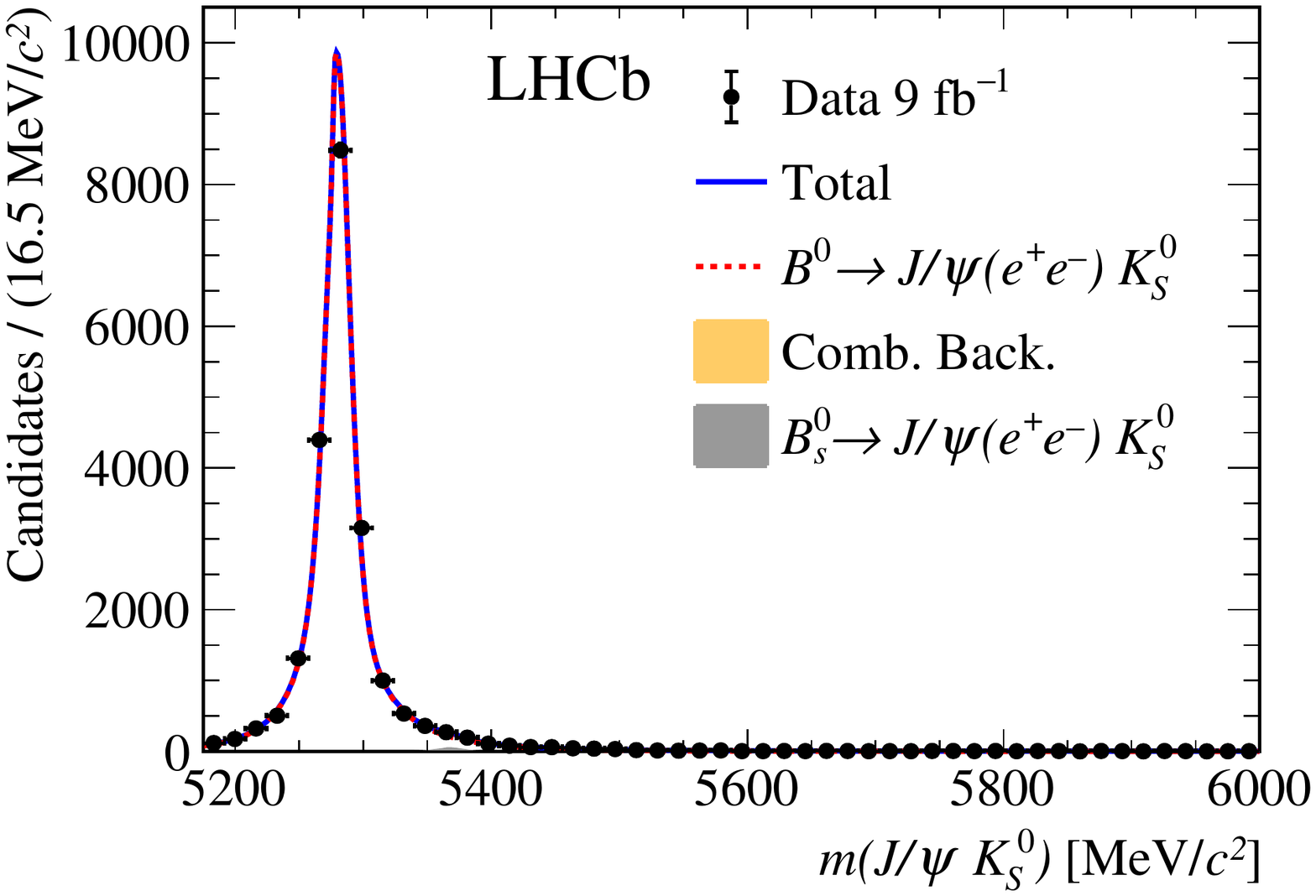}
     \includegraphics[width=0.48\linewidth]{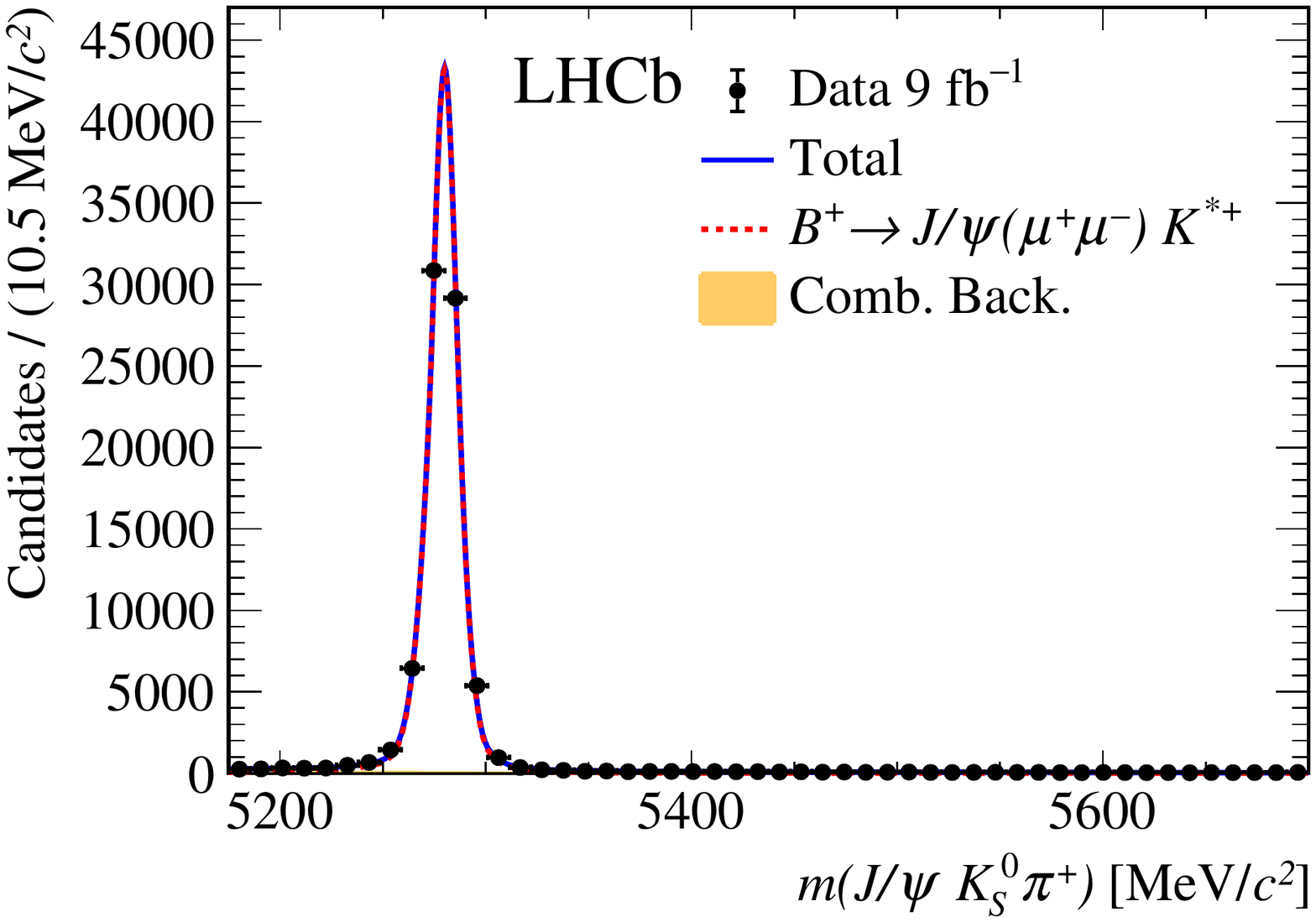}
     \includegraphics[width=0.48\linewidth]{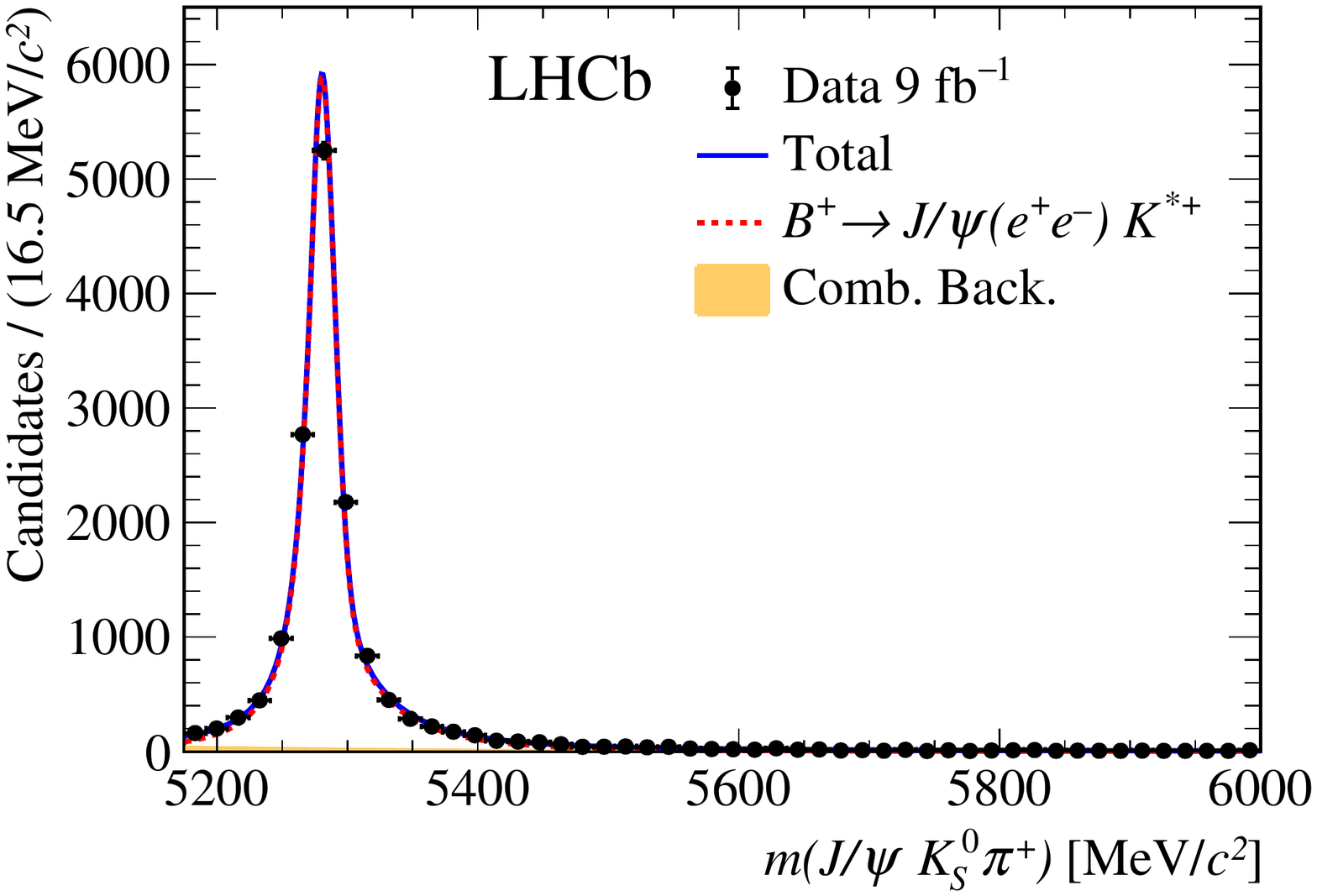}
\caption{\small Distributions of (top left)~$\jpsi(\mup\mun)\KS$ mass, (top right)~$\jpsi(\ep\en)\KS$ mass, (bottom left)~$\jpsi(\mup\mun)\KS\pip$ mass and (bottom right)~$\jpsi(\ep\en)\KS\pip$ mass with the fit models used to determine the control mode yields.}\label{fig:supplementary_controlfit_linscale}
\end{figure}

\begin{figure}[t]
    \centering
     \includegraphics[width=0.48\linewidth]{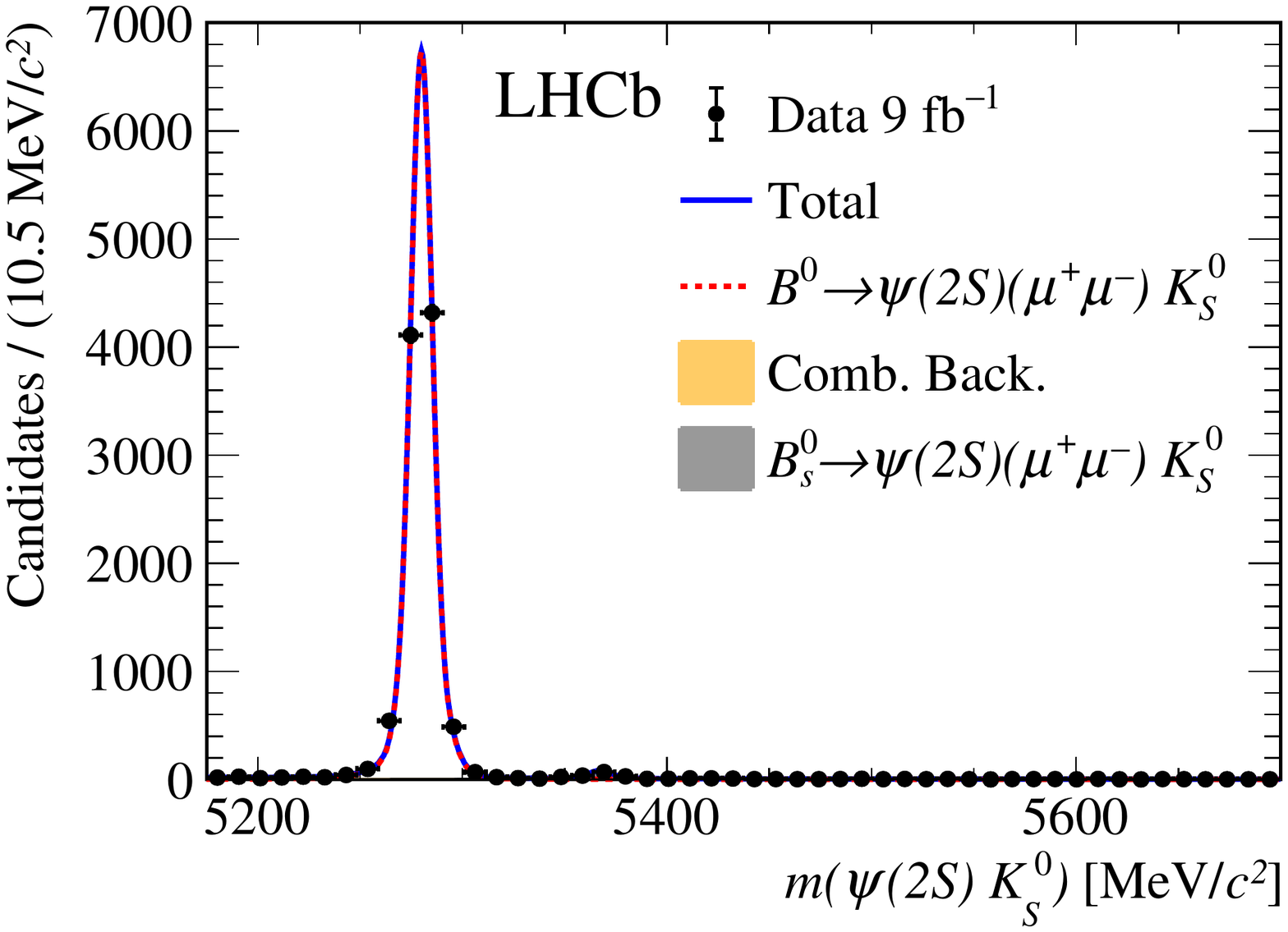}
     \includegraphics[width=0.48\linewidth]{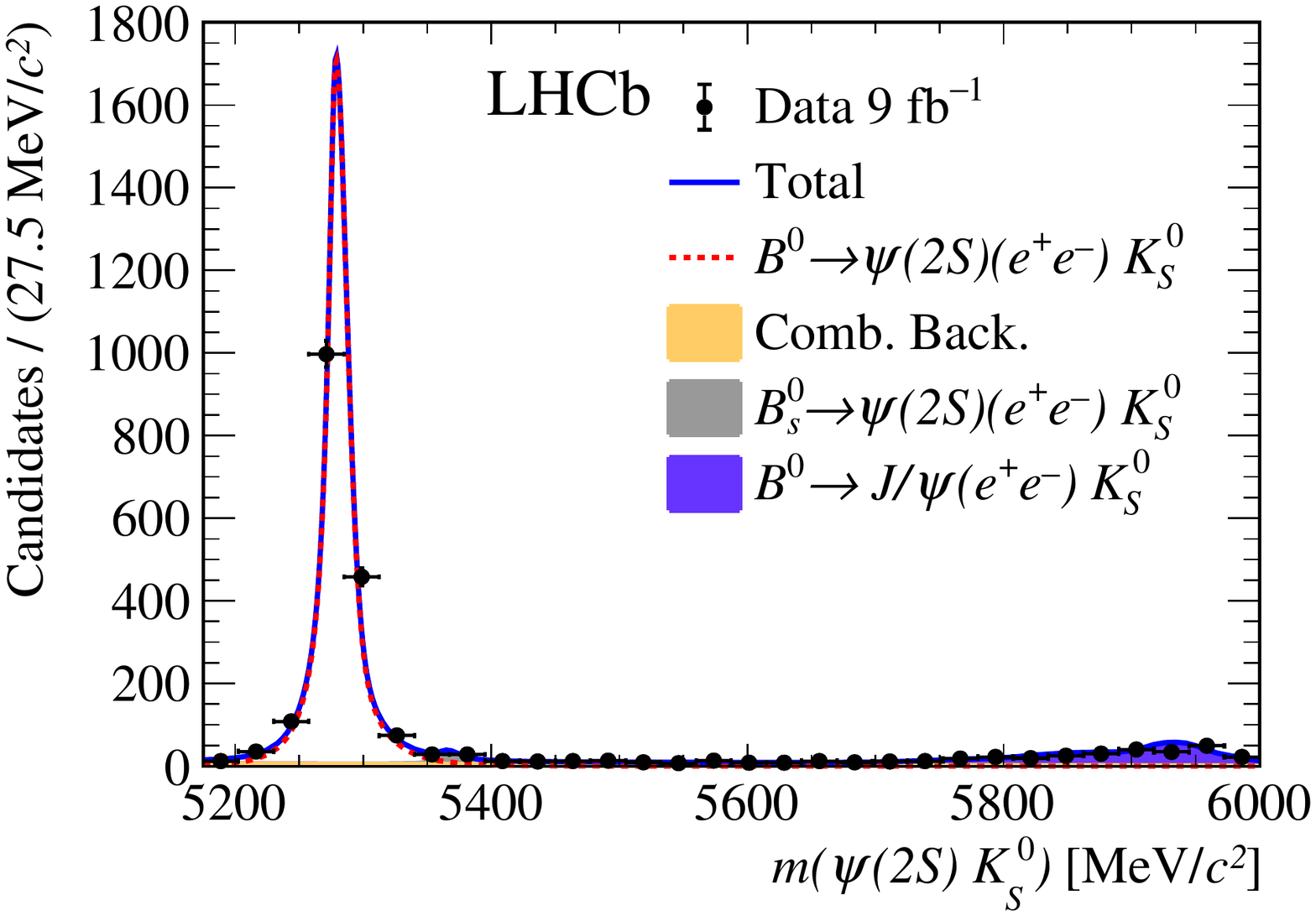}
     \includegraphics[width=0.48\linewidth]{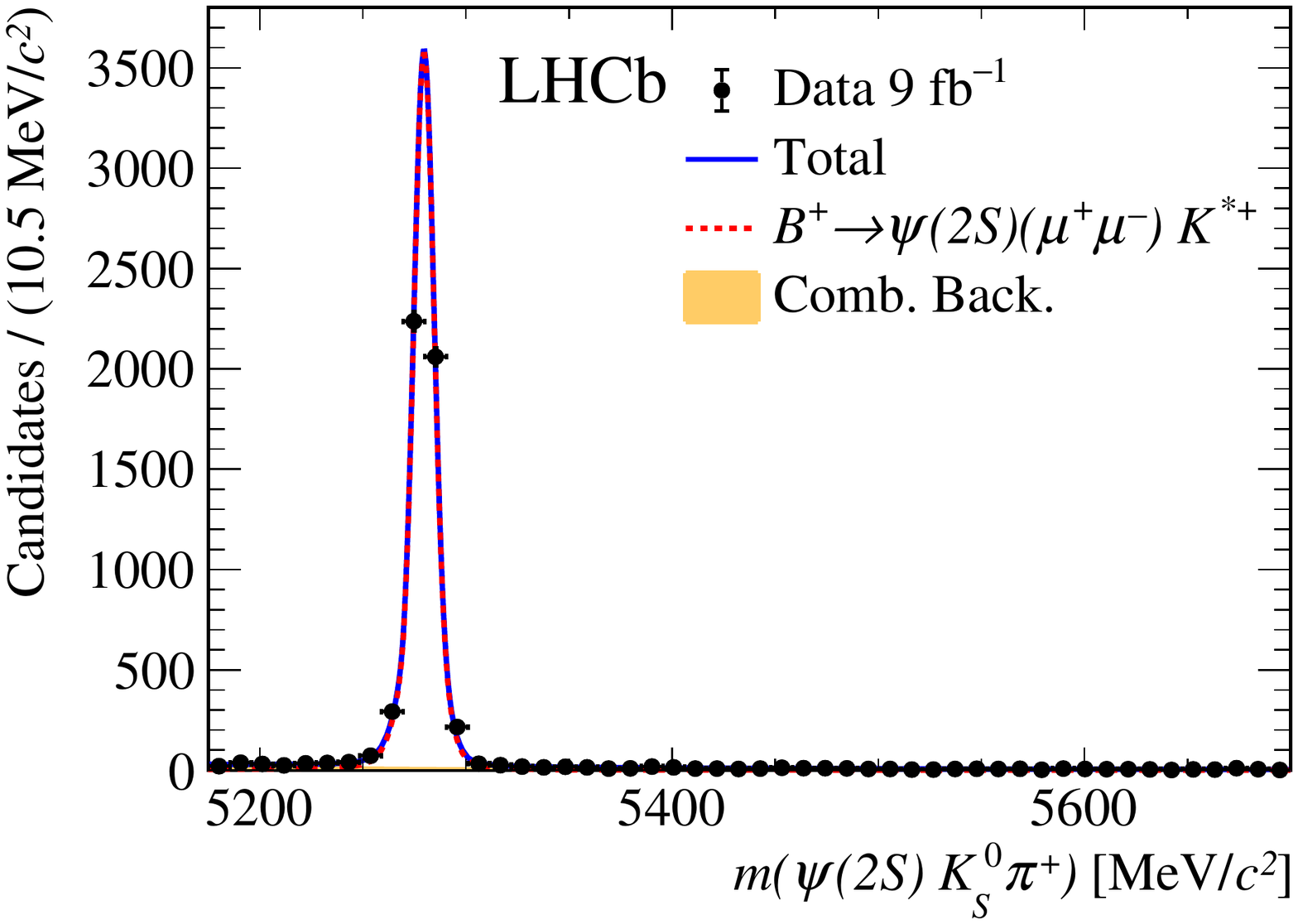}
     \includegraphics[width=0.48\linewidth]{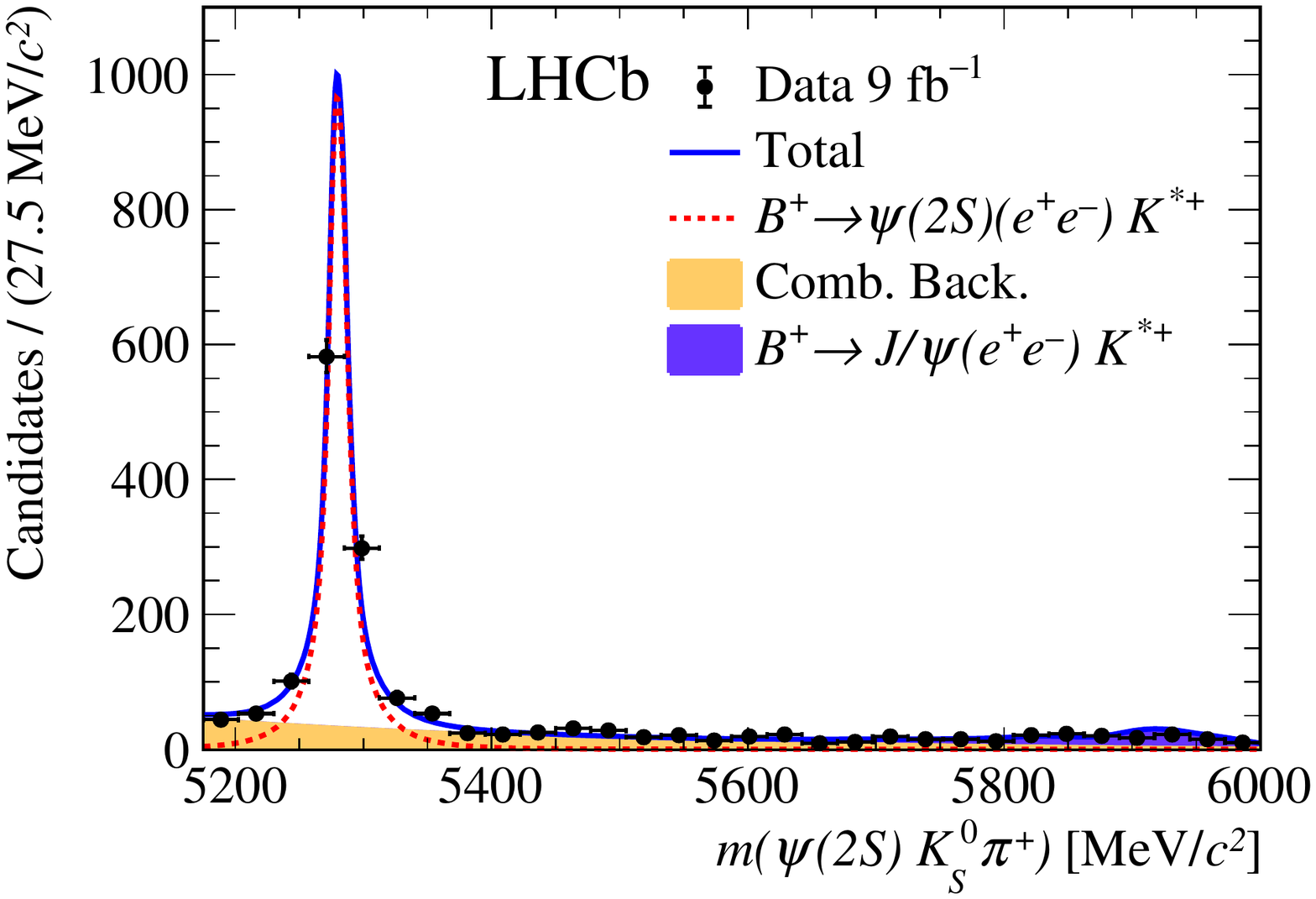}
\caption{\small 
Distributions of (top left)~$\psitwos(\mup\mun)\KS$ mass, (top right)~$\psitwos(\ep\en)\KS$ mass, (bottom left)~$\psitwos(\mup\mun)\KS\pip$ mass and (bottom right)~$\psitwos(\ep\en)\KS\pip$ mass with the fit models used to determine the \psitwos control mode yields.}\label{fig:supplementary_psitwosfit_linscale}
\end{figure}

\begin{figure}[t]
    \centering
     \includegraphics[width=0.48\linewidth]{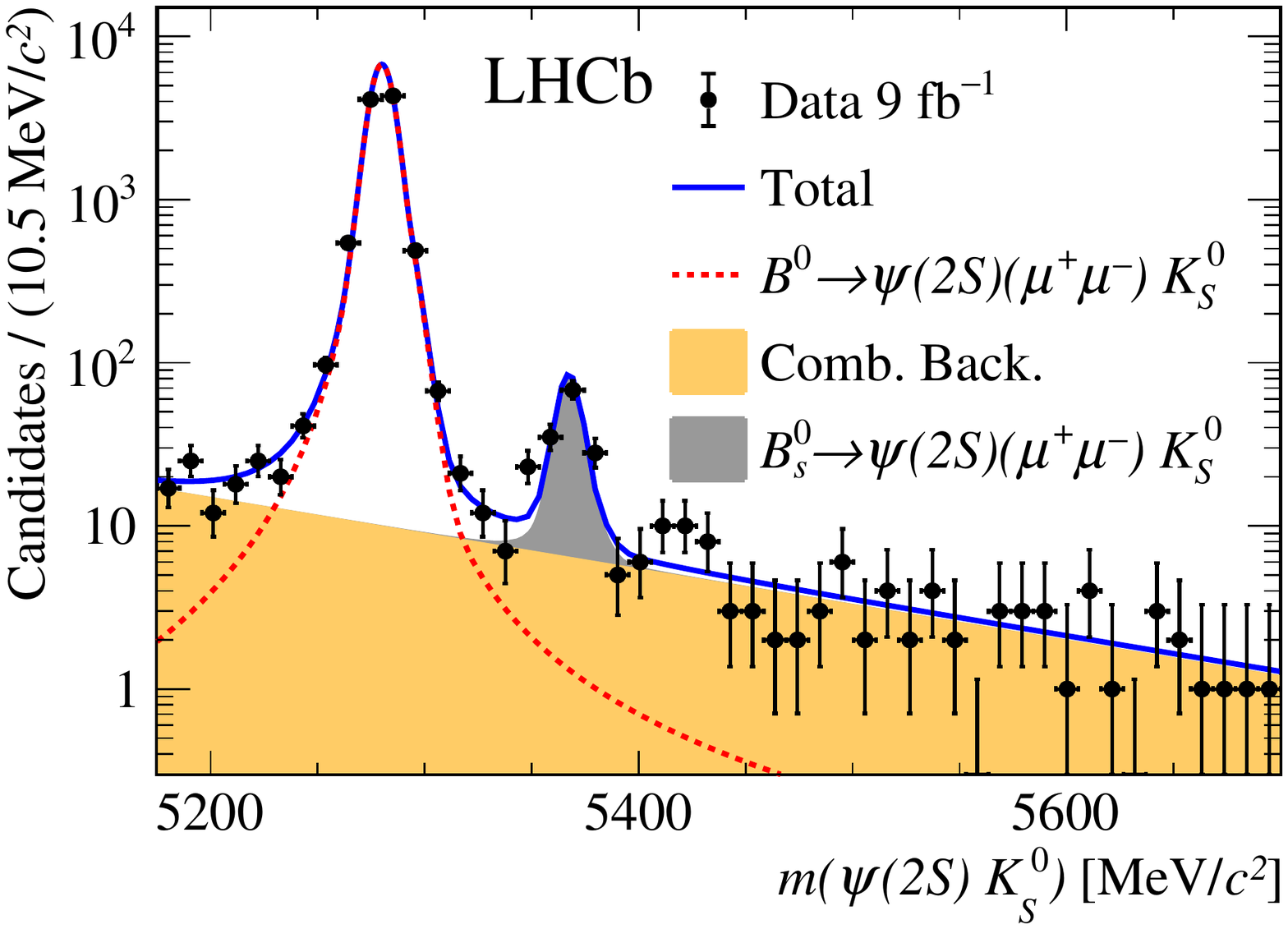}\hfill
     \includegraphics[width=0.48\linewidth]{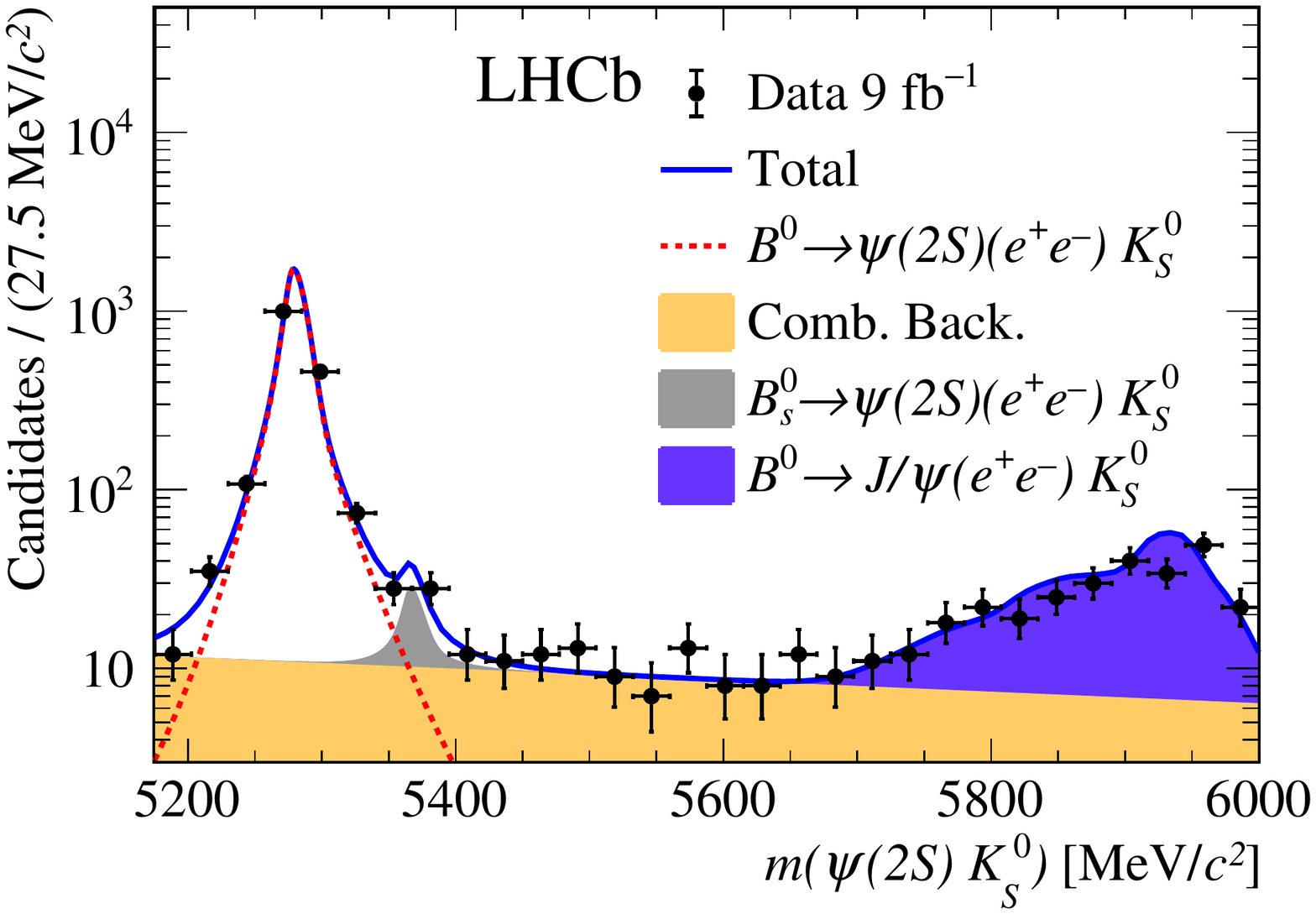}
     
     \includegraphics[width=0.48\linewidth]{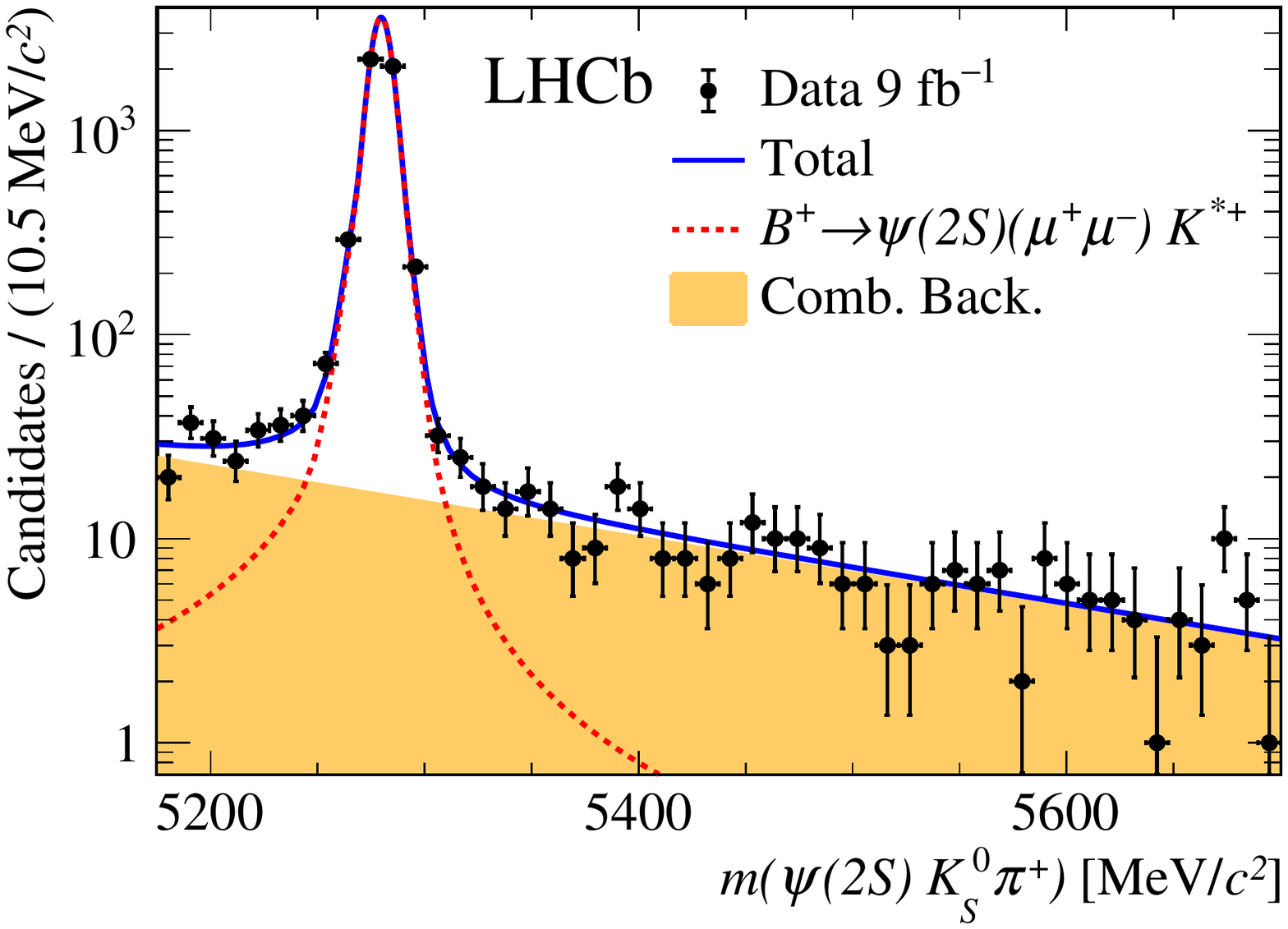}\hfill
     \includegraphics[width=0.48\linewidth]{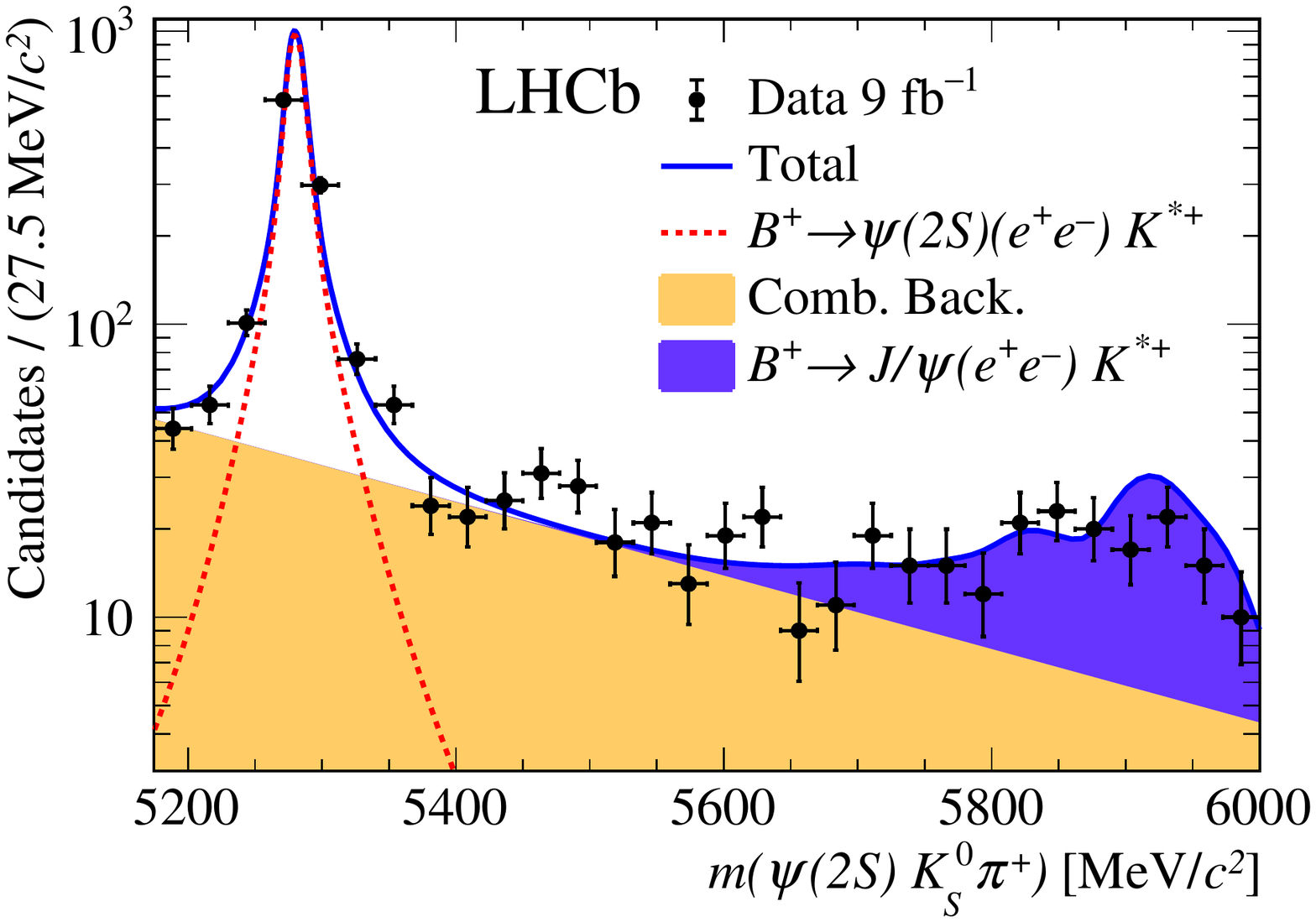}
\caption{\small 
Distributions of (top left)~$\psitwos(\mup\mun)\KS$ mass, (top right)~$\psitwos(\ep\en)\KS$ mass, (bottom left)~$\psitwos(\mup\mun)\KS\pip$ mass and (bottom right)~$\psitwos(\ep\en)\KS\pip$ mass with the fit models used to determine the \psitwos control mode yields.}\label{fig:supplementary_psitwosfit_logscale}
\end{figure}

This appendix contains supplemental material to the measurements. Eq. \ref{eqn:diffbfratios} gives the the \BdToKSee and \BuToKstee differential branching fractions divided by the branching fractions of their respective normalisation modes. Fig. \ref{fig:supplementary_eff_profiles} shows the total efficiencies for the four signal decay modes as function of \qsq. Figs. \ref{fig:supplementary_rjpsi_dilepton_opening_angle} and \ref{fig:supplementary_rjpsi_shell_clas} show  differential measurements of \rinvJpsiK. Figs. \ref{fig:supplementary_likelihood_scans_rx} and \ref{fig:supplementary_likelihood_scans_rare_ee_bfs} show the profile likelihood scans for \RinvK and the differential branching fractions. Fig. \ref{fig:supplementary_r_comparisons} compares the measurements of \RinvK presented in this Letter with previous measurements by the Belle collaboration. Fig. \ref{fig:supplementary_controlfit_linscale} shows the $\jpsi K^{(*)}$ invariant mass spectra and the maximum likelihood fits use to determine the \BToKJpsi control mode yields with a linear scale. Figs. \ref{fig:supplementary_psitwosfit_linscale} and \ref{fig:supplementary_psitwosfit_logscale} show the $\psitwos K^{(*)}$ invariant mass spectra and the maximum likelihood fits use to determine the \BToKpsitwos\ \,mode yields with linear and logarithmic scales.

In the main text, the differential branching fractions of \BdToKSee and \BuToKstee decays are reported. These are normalised using the world-average branching fractions for \BdToKzJpsiee\ and \BuToKstJpsiee\, respectively \cite{PDG2020}. The ratios of signal and control mode branching fractions used in this calculation are

\begin{align}
     \frac{\mathrm{d}\mathcal{B}\left(\BdToKzee \right)}{\mathrm{d}\qsq}\, \big/\, \mathcal{B}\left( \BdToKzJpsiee \right)& =\, \nonumber \\
     \left(4.9\,^{+1.2}_{-1.1} \left(\text{stat.}\right) \pm 0.2 \left(\text{syst.}\right) \right)& \times 10^{-4} \hspace{0.1cm} \gev^{-2}c^4, \nonumber \\
    \frac{\mathrm{d}\mathcal{B}\left(\BuToKstee \right)}{\mathrm{d}\qsq}\, \big/\, \mathcal{B}\left( \BuToKstJpsiee \right)& =\, \nonumber \\
    \left(1.08\,^{+0.22}_{-0.21} \left(\text{stat.}\right) \,^{+0.06}_{-0.04} \left(\text{syst.}\right) \right)& \times 10^{-3} \hspace{0.1cm} \gev^{-2}c^4,
    \label{eqn:diffbfratios}
\end{align}
where \BuToKstee candidates are selected with $m(\KS\pip)$ within 300 \mevcc of the world-average \Kstarp mass, whereas \BuToKstJpsiee candidates are selected in the range $792 < m(\KS\pip)/\mevcc < 992$.

\clearpage

% \input{appendix}

% % This should be taken out in the final paper
% \input{supplementary}

\addcontentsline{toc}{section}{References}
%\setboolean{inbibliography}{true}
\bibliographystyle{LHCb}
\bibliography{main,standard,LHCb-PAPER,LHCb-CONF,LHCb-DP,LHCb-TDR}

\newpage
% LHCb collaboration author list
% Data extracted on October 15th, 2021 at 12:56pm for paper reference LHCb-PAPER-2021-038
\centerline
{\large\bf LHCb collaboration}
\begin
{flushleft}
\small
R.~Aaij$^{32}$,
A.S.W.~Abdelmotteleb$^{56}$,
C.~Abell{\'a}n~Beteta$^{50}$,
F.~Abudin{\'e}n$^{56}$,
T.~Ackernley$^{60}$,
B.~Adeva$^{46}$,
M.~Adinolfi$^{54}$,
H.~Afsharnia$^{9}$,
C.~Agapopoulou$^{13}$,
C.A.~Aidala$^{87}$
% ...

{\footnotesize \it

% ...
$^{9}$Universit{\'e} Clermont Auvergne, CNRS/IN2P3, LPC, Clermont-Ferrand, France\\
% ...
$^{13}$LPNHE, Sorbonne Universit{\'e}, Paris Diderot Sorbonne Paris Cit{\'e}, CNRS/IN2P3, Paris, France\\
% ...
$^{32}$Nikhef National Institute for Subatomic Physics, Amsterdam, Netherlands\\
% ...
$^{46}$Instituto Galego de F{\'\i}sica de Altas Enerx{\'\i}as (IGFAE), Universidade de Santiago de Compostela, Santiago de Compostela, Spain\\
% ...
$^{50}$Physik-Institut, Universit{\"a}t Z{\"u}rich, Z{\"u}rich, Switzerland\\
% ...
$^{54}$H.H. Wills Physics Laboratory, University of Bristol, Bristol, United Kingdom\\
% ...
$^{56}$Department of Physics, University of Warwick, Coventry, United Kingdom\\
% ...
$^{60}$Oliver Lodge Laboratory, University of Liverpool, Liverpool, United Kingdom\\
% ...
$^{68}$Syracuse University, Syracuse, NY, United States\\
% ...
$^{87}$University of Michigan, Ann Arbor, United States, associated to $^{68}$\\
\bigskip
% ...
}
\end{flushleft}

% \clearpage
% \input{prl-justification}

\end{document}